\documentclass[twocolumn,aps,pra,longbibliography,showpacs,superscriptaddress,nofootinbib,10pt]{revtex4-1}
\usepackage[latin1]{inputenc}
\usepackage{graphicx,dcolumn,bm}
\usepackage{subfigure}
\usepackage{color}
\usepackage[colorlinks,hyperindex]{hyperref}
\usepackage{amsfonts}
\usepackage{amssymb}
\usepackage{amsmath}
\usepackage{latexsym}
\usepackage{times,txfonts}
\usepackage{epstopdf}

\usepackage{mathrsfs}
\usepackage{arydshln}
\usepackage[sans]{dsfont}

\usepackage{epsfig}

\definecolor{Blue}{rgb}{0.0,0.0,1}
\definecolor{Red}{rgb}{1,0.0,0.0}
\definecolor{Green}{rgb}{0,0.5,0.0}

\begin{document}

\title{Generalized Geometric Quantum Speed Limits}
\author{Diego Paiva Pires}
\email{diegopaivapires@gmail.com}
\thanks{\newline The first two authors contributed equally to this work.}
\affiliation{Instituto de F\'{i}sica de S\~{a}o Carlos,
Universidade de S\~{a}o Paulo, CP 369, 13560-970, S\~{a}o Carlos, SP, Brasil}
\author{Marco Cianciaruso}
\email{cianciaruso.marco@gmail.com}
\thanks{\newline The first two authors contributed equally to this work.}
\affiliation{School of Mathematical Sciences, The University of Nottingham, University Park, Nottingham NG7 2RD, United Kingdom}
\affiliation{Dipartimento di Fisica ``E. R. Caianiello'', Universit\`{a} degli Studi di Salerno,
Via Giovanni Paolo II, I-84084 Fisciano (FA), Italy}
\affiliation{INFN, Sezione di Napoli, Gruppo Collegato di Salerno, I-84084 Fisciano (FA), Italy}
\author{Lucas C. C\'{e}leri}
\email{lucas@chibebe.org}
\affiliation{Instituto de F\'{i}sica, Universidade Federal de Goi\'{a}s, 74.001-970 Goi\^{a}nia, Goi\'{a}s, Brazil}
\author{Gerardo Adesso}
\email{gerardo.adesso@nottingham.ac.uk}
\affiliation{School of Mathematical Sciences, The University of Nottingham,
University Park, Nottingham NG7 2RD, United Kingdom}
\author{Diogo O. Soares-Pinto}
\email{dosp@ifsc.usp.br}
\affiliation{Instituto de F\'{i}sica de S\~{a}o Carlos,
Universidade de S\~{a}o Paulo, CP 369, 13560-970, S\~{a}o Carlos, SP, Brasil}
\begin{abstract}
The attempt to gain a theoretical understanding of the concept of time in quantum mechanics has triggered significant progress towards the search for faster and more efficient quantum technologies. One of such advances consists in the interpretation of the time-energy uncertainty relations as lower bounds for the minimal evolution time between two distinguishable states of a quantum system, also known as quantum speed limits. We investigate how the non uniqueness of a bona fide measure of distinguishability defined on the quantum state space affects the quantum speed limits and can be exploited in order to derive improved bounds. Specifically, we establish an infinite family of quantum speed limits valid for unitary and nonunitary evolutions, based on an elegant information geometric formalism. Our work unifies and generalizes existing results on quantum speed limits, and provides instances of novel bounds which are tighter than any established one based on the conventional quantum Fisher information. We illustrate our findings with relevant examples, demonstrating the importance of choosing different information metrics for open system dynamics, as well as clarifying the roles of classical populations versus quantum coherences, in the determination and saturation of the speed limits. Our results can find applications in the optimization and control of quantum technologies such as quantum computation and metrology, and might provide new insights in fundamental investigations of quantum thermodynamics.
\end{abstract}


\maketitle


\section{Introduction}
\label{sec:introductionI}

Quantum mechanics relies on counterintuitive features which challenge our merely classical perception of Nature. One of the most fundamental quantum aspects lies in the impossibility of knowing simultaneously and with certainty two incompatible properties of a quantum system~\cite{1927_ZPhys_43,*1929_PhysRev_163,*PhysRevLett.111.230401}. Contrarily to the well understood uncertainty relation between any two non-commuting observables, the time-energy uncertainty relation still represents a controversial issue~\cite{WPauli_23_278,*WPauli_24_272,*Bohr_Nature_121_580}, although the last decades witnessed several attempts towards its explanation~\cite{1961_PhysRev_122_1649,*1974_IntJTheorPhys_11_357,*1978_IntJTheorPhys_13_379}. This effort led to the interpretation of the time-energy uncertainty relation as a so-called {\it quantum speed limit} (QSL), i.e. the ultimate bound imposed by quantum mechanics on the minimal evolution time between two distinguishable states of a system~\cite{1945_JPhysURSS_9_249,1973_INCimentoA_16_232_Fleming,1983_JPhysAMathGen_16_2993,1990_PhysRevLett_65_1697,1991_PhysLettA_159_105,uhlmann1992energy,
1992_AmJPhys_60_182_Vaidman,1993_AmJPhys_61_935_Uffink,1995_PhysRevA_52_2576,1997_PhysLettA_231_29,1999_PhysLettA_262_296,2003_36_5587_JPhysAMathGen_Dorje_Brody,1992_PhysicaD_120_188,
2009_PhysRevLett_103_160502,2003_PhysRevA_67_052109,2010_PhysRevA_82_022107,2013_JMathPhys_46_335302,
1993_PhysRevLett_70_3365,RevModPhys.67.759,PhysRevLett.111.260501,PhysRevA.90.032110,2004_PhysicaD_1_189_Luo,2005_JPhysAMathGen_Luo,
PhysRevA.90.012303_Russell,2014_JPAMathPhys_47_215301, PhysRevA.91.042330,1506.03199_Mondal_Datta_Sk,2013_PhysRevLett_110_050402,2013_PhysRevLett_110_050403,2013_PhysRevLett_111_010402,2014_arXiv_14081227v1_Uzdin,2014_NatPhys_4_4890,PhysRevA.91.022102_LiuXuZhu,PhysRevA.89.012307,2015_JPhysAMathTheor_48_045301,arXiv:1403.5182v2,PhysRevLett.102.017204_Lieb_Robinson,arXiv:1505.07850}. QSLs have been widely investigated within the quantum information setting, since their understanding offers a route to design faster and optimized information processing devices~\cite{2009_PhysRevLett_103_240501,*2014_JPhysBAtMolOptPhys_Deffner_47_145502}, thus attracting constant interest in quantum optimal control, quantum metrology~\cite{NatPhoton_5_222_GSLM}, quantum computation and communication~\cite{PhysRevA.74.030303}. Interestingly, it has been recently recognized that QSLs play a fundamental role also in quantum thermodynamics~\cite{2010_PhysRevLett_105_170402}.

In a seminal work, Mandelstam and Tamm (MT)~\cite{1945_JPhysURSS_9_249} reported a QSL for a quantum system that evolves between two distinguishable pure states, $|\psi(0)\rangle$ and $|\psi(\tau)\rangle$, via a unitary dynamics generated by a time independent Hamiltonian $H$. The ensuing lower bound on the evolution time is given by $\tau \geq \hbar \arccos(|\langle\psi(\tau)|\psi(0)\rangle|)/\Delta E$, where $(\Delta E)^2 = {\langle{H^2}\rangle - {\langle{H}\rangle^2}}$ is the variance of the energy of the system with respect to the initial state. Several years later, Anandan and Aharanov~\cite{1990_PhysRevLett_65_1697} extended the MT bound to time dependent Hamiltonians by using a geometric approach which exploits the Fubini-Study metric defined on the space of quantum pure states. Specifically, they simply used the fact that the geodesic length between two distinguishable pure states according to the Fubini-Study metric, i.e. their Bures angle, is a lower bound to the length of any path connecting the same states. Over half a century after the MT result,
Margolus and Levitin (ML)~\cite{1992_PhysicaD_120_188} provided a different QSL on the time evolution of a closed system whose Hamiltonian is time independent and evolving between two orthogonal pure states. This bound reads as $\tau \geq \pi\hbar/(2E)$, where $E = \langle H\rangle$ is the mean energy. Although the ML bound is tight, it does not recover the MT one whatsoever. Therefore, the quantum speed limit for unitary dynamics, when restricting to orthogonal pure states, can be made tighter by combining these two independent results as $\tau \geq \max\{\pi\hbar/(2\Delta E),\pi\hbar/(2E)\}$~\cite{2009_PhysRevLett_103_160502}.

All these results attracted a considerable interest in the subject. Giovannetti \textit{et al.}~\cite{2003_PhysRevA_67_052109} extended the ML QSL to the case of arbitrary mixed states and also showed that entanglement can speed up the dynamical evolution of a closed composite system.
A plethora of other extensions and applications of QSLs for unitary processes has been investigated in Refs.~\cite{1973_INCimentoA_16_232_Fleming,1983_JPhysAMathGen_16_2993,1991_PhysLettA_159_105,uhlmann1992energy,1992_AmJPhys_60_182_Vaidman,1993_AmJPhys_61_935_Uffink,1995_PhysRevA_52_2576, 1997_PhysLettA_231_29, 1999_PhysLettA_262_296,2003_36_5587_JPhysAMathGen_Dorje_Brody,2010_PhysRevA_82_022107,2013_JMathPhys_46_335302,1993_PhysRevLett_70_3365,RevModPhys.67.759,PhysRevLett.111.260501,PhysRevA.90.032110,2004_PhysicaD_1_189_Luo,
2005_JPhysAMathGen_Luo,PhysRevA.90.012303_Russell,1506.03199_Mondal_Datta_Sk,2014_JPAMathPhys_47_215301,PhysRevA.91.042330}. For example, in Ref.~\cite{PhysRevA.91.042330} some of us have recently shown that the rate of change of the distinguishability between the initial and the evolved state of a closed quantum system can provide a lower bound for an indicator of quantum coherence based on the Wigner-Yanase information between the evolved state and the Hamiltonian generating the evolution.

Since any information processing device is inevitably subject to environmental noise, QSLs have been also investigated in the nonunitary realm. Taddei \textit{et al.}~\cite{2013_PhysRevLett_110_050402} and del Campo {\it et al}.~\cite{2013_PhysRevLett_110_050403} were the first to extend the MT bound to any physical process, being it unitary or not. Specifically, Ref.~\cite{2013_PhysRevLett_110_050402} exploits the quantum Fisher information metric on the whole quantum state space and represents a natural extension of the idea used in Ref.~\cite{1990_PhysRevLett_65_1697}, whereas Ref.~\cite{2013_PhysRevLett_110_050403} exploits the relative purity. Then, Deffner and Lutz~\cite{2013_PhysRevLett_111_010402} extended the ML bound to open quantum systems by adopting again a geometric approach using the Bures angle. These authors have also introduced a new sort of bound, which is tighter than both the ML and MT ones, and shown that non-Markovianity can speed up the quantum evolution. Some other works have then provided a QSL for open system dynamics by using the relative purity, whose usefulness ranges from thermalization phenomena~\cite{2014_arXiv_14081227v1_Uzdin} to the relativistic effects on the QSL~\cite{2014_NatPhys_4_4890}. Further developments include the role of entanglement in QSL for open dynamics~\cite{PhysRevA.91.022102_LiuXuZhu,PhysRevA.89.012307,2015_JPhysAMathTheor_48_045301}, QSL in the one-dimensional perfect quantum state transfer~\cite{PhysRevA.74.030303}, and the experimental realizability of measuring QSLs through interferometry devices~\cite{arXiv:1403.5182v2}. Finally, a subtle connection was recently highlighted between QSLs and the maximum interaction speed in quantum spin systems~\cite{PhysRevLett.102.017204_Lieb_Robinson}, with implications for quantum error correction and the relaxation time of many-body systems~\cite{arXiv:1505.07850}.

Distinguishing between two states of a system being described by a probabilistic model stands as the paradigmatic task of information theory. Information geometry, in particular, applies methods of differential geometry in order to achieve this task~\cite{amari_infor_geom}. The set of states of both classical and quantum systems is indeed a Riemannian manifold, that is the set of probability distributions over the system phase space and the set of density operators over the system Hilbert space, respectively. Therefore it seems natural to use any of the possible Riemannian metrics defined on such sets of states in order to distinguish any two of its points. However, it is also natural to assume that for a Riemannian metric to be bona fide in quantifying the distinguishability between two states, it must be contractive under the physical maps that represent the mathematical counterpart of noise, i.e. stochastic maps in the classical settings and completely positive trace preserving maps in quantum. Interestingly, \v{C}encov's theorem states that the Fisher information metric is the only Riemannian metric on the set of probability distributions that is contractive under stochastic maps~\cite{Chencov_book}, thus leaving us with only one choice of bona fide Riemannian geometric measure of distinguishability within the classical setting. On the contrary, it turns out that the quantum Fisher information metric~\cite{1981_PhysRevLett_23_357,1994_PhysRevLett_72_3439} is not the only contractive Riemannian metric on the set of density operators, but rather there exists an infinite family of such metrics~\cite{2003_JMathPhys_44_3752}, as characterized by the Morozova, \v{C}encov and Petz  theorem~\cite{1990_Itogi_Nauki_i_Tehniki_36_69,
1996_LettMathPhys_38_221,*2002_JPhysA_35_929,*Petz_Hiai_2009_430_3105,*2013_PhysRevA_87_032324}.

\begin{figure}[t]
\centering
\includegraphics[width=7.5cm]{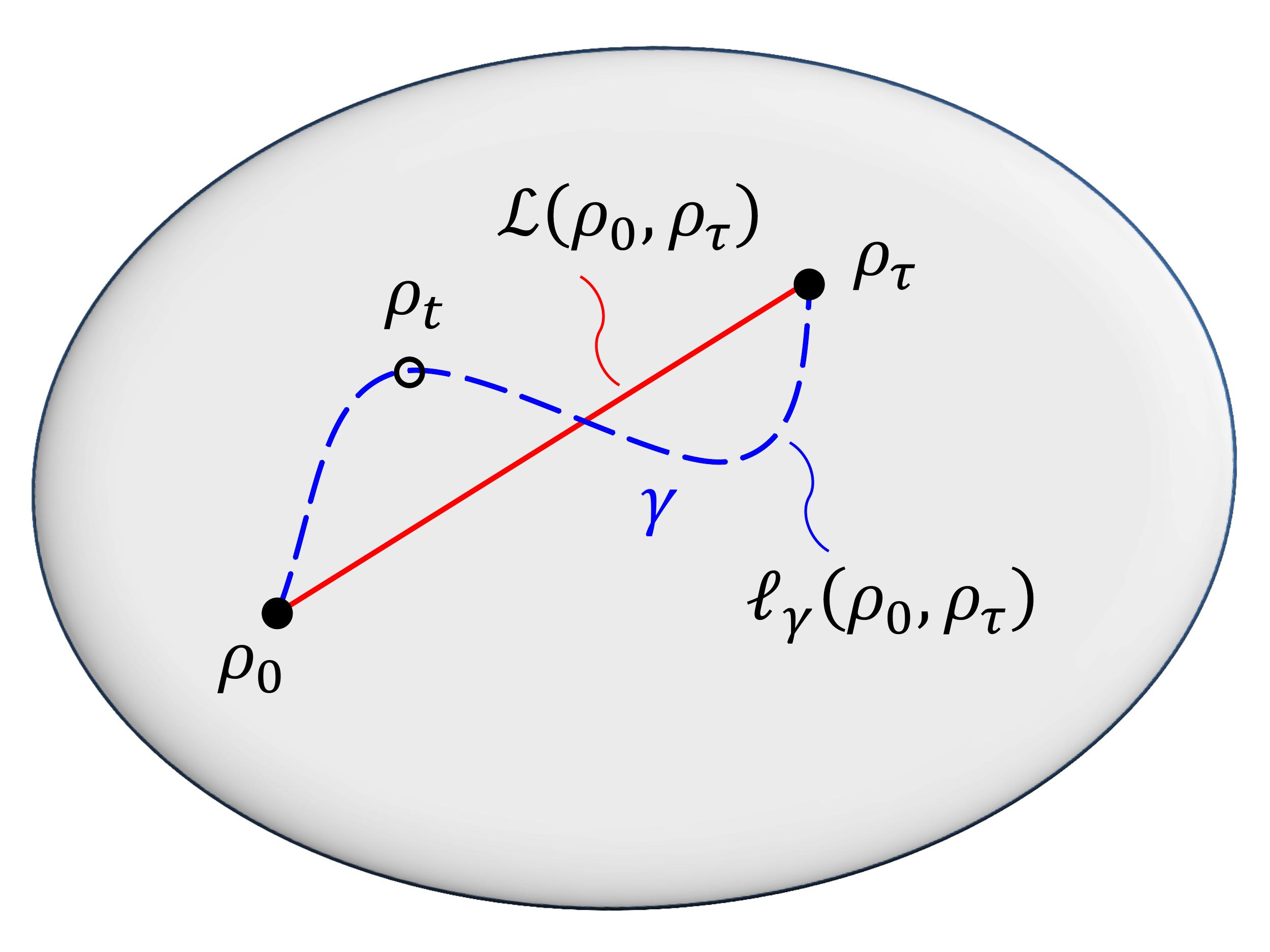}
\caption{(Color online). Illustration of geometric quantum speed limits. The dashed blue curve is the path $\gamma$ in the quantum state space representing a generic evolution between an initial state $\rho_0$ and a final state $\rho_\tau$, parameterized by time $t \in [0,\tau]$. Given a metric on the quantum state space, the length of this path is denoted by $\ell_\gamma(\rho_0,\rho_\tau)$. The solid red curve denotes the geodesic connecting $\rho_0$ to $\rho_\tau$, whose length is given by $\mathcal{L}(\rho_0,\rho_\tau)$. Quantum speed limits originate from the fact that the geodesic amounts to the path of shortest length among all physical evolutions between the given initial and final states: $\mathcal{L}(\rho_0,\rho_\tau) \leq \ell_\gamma(\rho_0,\rho_\tau)$ $\forall \gamma$. Such inequality can be interpreted as follows. For any given physical evolution $\gamma$ from $\rho_0$ to $\rho_\tau$, and according to any valid metric, the maximum distance between the initial $\rho_0$ and the final state $\rho_\tau$ is the length of the path $ \ell_\gamma(\rho_0,\rho_\tau)$ followed by the system. The ensuing minimal time necessary for this distance to reach a chosen value is the time at which the path length reaches this value. This interpretation provides a neat criterion for the saturation of the lower bound on the evolution time, that is when the dynamical evolution coincides with a geodesic of the considered metric. Here we establish a general family of geometric quantum speed limits with respect to an infinite hierarchy of contractive Riemannian metrics on the space of quantum states, unifying and extending previous results under an  information geometry framework.
}\label{figQS}
\end{figure}

In this paper, we construct a new fundamental family of geometric QSLs (see Fig.~\ref{figQS}) which is in one to one correspondence with the family of contractive Riemannian metrics characterized by the Morozova, \v{C}encov and Petz theorem.  We demonstrate how such non uniqueness of a bona fide measure of distinguishability defined on the quantum state space affects the QSLs and can be exploited in order to look for tighter bounds. Our approach is versatile enough to provide a unified picture, encompassing both unitary and nonunitary dynamics, and is easy to handle, requiring solely the spectral decomposition of the evolved state. This family of bounds is naturally tailored to the general case of initial mixed states and clearly separates the contribution of the populations of the evolved state and the coherences of its time variation, thus clarifying their individual role in driving the evolution.

We formulate in rigorous terms the problem of identifying the tightest bound within our family for any given dynamics. While such a problem is unfeasibly hard to address in general, we establish concrete steps towards its solution in practical scenarios. Specifically, we show explicit instances of QSLs which make use of some particular contractive Riemannian metric such as the Wigner-Yanase skew information and can be provably tighter than the corresponding QSLs obtained with the conventional quantum Fisher information. These instances are relevant in metrological settings. Overall this work provides one of the most comprehensive and powerful approaches to QSLs, with potential impact on the characterization and control of quantum technologies.

The paper is organized as follows. In Sec.~\ref{sec:sectionII} we review the relation between statistical distinguishability and the contractive Riemannian metrics on the quantum state space characterized by the Morozova, \v{C}encov and Petz theorem. Section~\ref{sec:sectionIII} provides a new generalized geometric derivation of a family of QSLs which is in one to one correspondence with the family of such metrics. In Sec.~\ref{sec:examplesPD_AD} we illustrate and compare the obtained bounds for both unitary and nonunitary evolutions. Finally, in Sec.~\ref{sec:sectionIV} we present our conclusions.


\section{Geometric measures of distinguishability}
\label{sec:sectionII}

According to the standard formulation of quantum mechanics, any quantum system is associated with a Hilbert space $\mathcal{H}$ and its states are represented by the Riemannian manifold $\mathcal{S}=\mathcal{D}(\mathcal{H})$ of density operators over $\mathcal{H}$, i.e. the set of positive semi-definite and trace one operators over the carrier Hilbert space. A Riemannian metric over $\mathcal{S}$ is said to be contractive if the corresponding geodesic distance $\mathcal{L}$ contracts under physical maps, which means satisfy the inequality $\mathcal{L}(\Lambda(\rho),\Lambda(\sigma))\leq \mathcal{L}(\rho,\sigma)$ for any completely positive trace preserving map $\Lambda$ and any $\rho,\sigma\in\mathcal{S}$. The Morozova, \v{C}encov and Petz theorem provides us with a characterization of such metrics in the finite-dimensional case, by constructing a one to one correspondence between them and the Morozova-\v{C}encov (MC) functions, a function $f(t):\mathbb{R}_+\rightarrow \mathbb{R}_+$ which is ($i$) operator monotone: for any semi-positive definite operators $A$ and $B$ such that $A \leq B$, then $f(A) \leq f(B)$; ($ii$) self-inversive: it fulfils the functional equation $f(t) = tf(1/t)$; and ($iii$) normalized: $f(1) = 1$. Specifically, the Morozova, \v{C}encov and Petz theorem states that every contractive Riemannian metric $\textbf{g}^f$ assigns, up to a constant factor, the following squared infinitesimal length between two neighboring density operators $\rho$ and $\rho+d\rho$~\cite{1996_LinAlgApl_244_81}
\begin{equation}\label{eq:contractivesquaredinfinitesimallength}
d{s^2} := {\textbf{g}^f_\rho}(d\rho,d\rho),
\end{equation}
with
\begin{equation}
 \label{eq:cencovmorozovfunc001}
{\textbf{g}^f_\rho}(A,B) = \frac{1}{4} \text{Tr}[A \, {c^f}(\textbf{L}_\rho,\textbf{R}_\rho) \, B] ~,
\end{equation}
where $A$ and $B$ are any two traceless hermitian operators, and
\begin{equation}
 \label{eq:cencovmorozovfunc002}
{c^f}(x,y) := \frac{1}{yf(x/y)}
\end{equation}
is a symmetric function, ${c^f}(x,y) = {c^f}(y,x)$, which fulfills ${c^f}(\alpha x,\alpha y) = {\alpha^{-1}}{c^f}(x,y)$, with $f(t)$ being a MC function, and finally $\textbf{L}_\rho, \textbf{R}_\rho : \mathcal{B}(\mathcal{H})\rightarrow \mathcal{B}(\mathcal{H})$ are two linear superoperators defined on the set $\mathcal{B}(\mathcal{H})$ of linear operators over $\mathcal{H}$ as follows: $\textbf{L}_\rho A = \rho A$ and $\textbf{R}_\rho A = A \rho$. We stress again that each contractive Riemannian metric is arbitrary up to a constant factor. In accordance with Ref.~\cite{Ingemar_Bengtsson_Zyczkowski}, we have chosen the factor $1/4$ in order to make the entire family of contractive Riemannian metrics collapse to the classical Fisher information metric when $\rho$ and $d\rho$ commute.

In order to make Eq.~\eqref{eq:contractivesquaredinfinitesimallength} more explicit, we can write the density operator $\rho$ in its spectral decomposition, $\rho = {\sum_j}{p_j}|{j}\rangle\langle{j} \, |$, with $0 < {p_j} \leq 1$ and ${\sum_j}{p_j} = 1$, and get~\cite{Ingemar_Bengtsson_Zyczkowski}
\begin{equation}
 \label{eq:mrzcncpetz0001}
{ds^2} = \frac{1}{4} \left[{\sum_{j}}\frac{(d\rho_{jj})^2}{p_j} + {2}{\sum_{j < l}} \, {c^f}({p_j},{p_l}){|{d\rho_{jl}}|^2}\right]~,
\end{equation}
where ${d\rho_{jl}}:= \langle{j}|d\rho|{l}\rangle$ and we note that the summation is constrained to the requirement ${p_j} > 0$. Equation~(\ref{eq:mrzcncpetz0001}) is crucial since it clearly identifies two separate contributions to any contractive Riemannian metric. The first term, which is common to all the family, depends only on the populations $p_j$ of $\rho$ and can be seen as the classical Fisher information metric at the probability distribution $p_j$. The second term, which is responsible for the non uniqueness of a contractive Riemmanian metric on the quantum state space, is instead only due to the coherences of $d\rho$ with respect to the eigenbasis of $\rho$ and is a purely quantum contribution expressing the non-commutativity between the operators $\rho$ and $\rho + d\rho$. Finally, for all the contractive Riemannian metrics that can be naturally extended to the boundary of pure states, such that $f(0)\neq 0$, the Fubini-Study metric appears always to be such extension up to a constant factor, so that the non uniqueness of a contractive Riemannian geometry can be only witnessed when considering quantum mixed states. This is the reason for which only the mixed states will be relevant in our analysis, whose aim is exactly to investigate the freedom in the choice of several inequivalent bona fide measures of indistinguishability in order to get tighter QSLs.

\begin{figure}[t]
\centering
\includegraphics[scale=0.45]{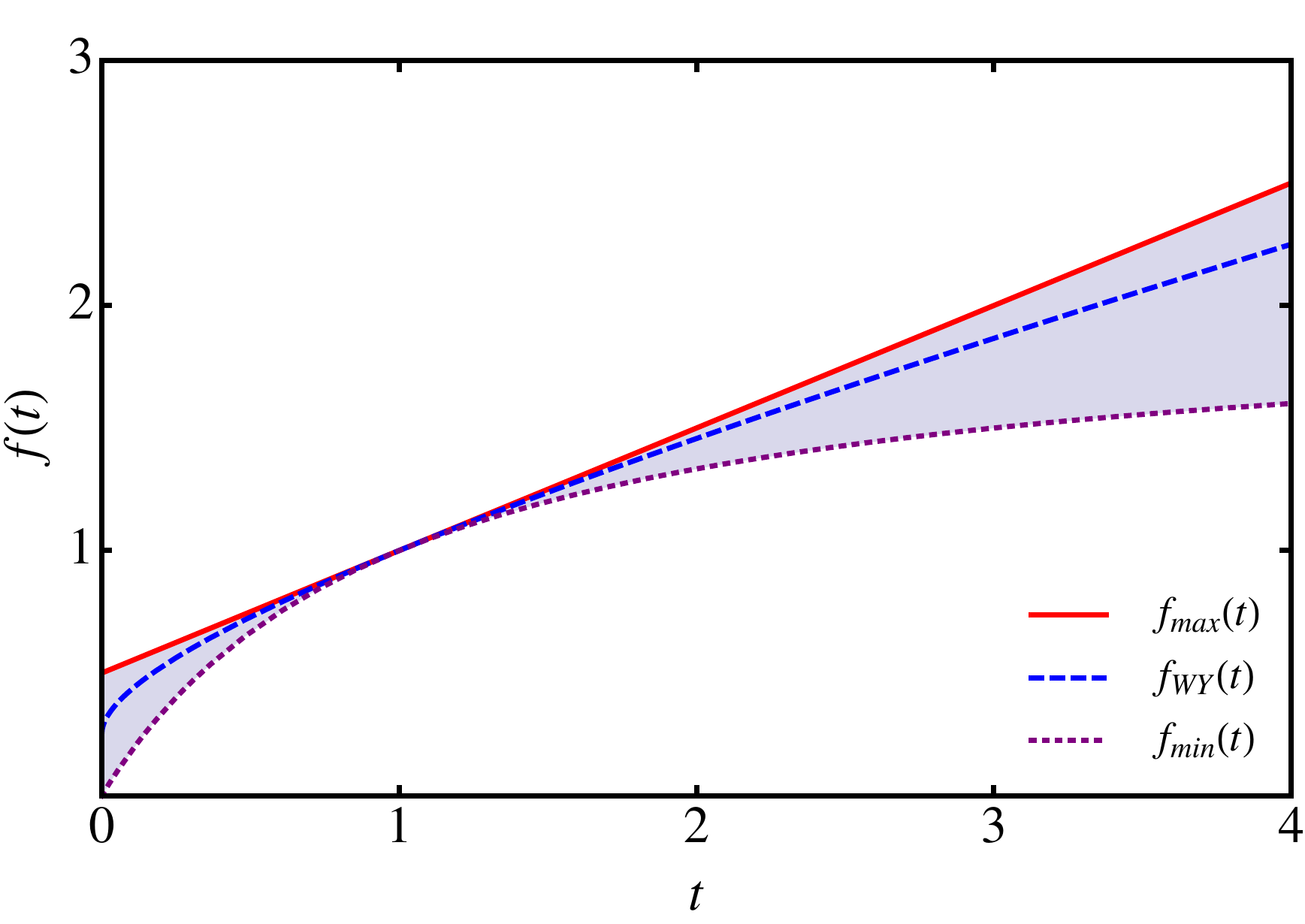}
\caption{(Color online). The family of Morozova-\v{C}encov (MC) functions is upper bounded by a maximal one, ${f_{max}}(t) = (1 + t)/2$, corresponding to the quantum Fisher information metric (red solid line), and lower bounded by a minimal one, ${f_{min}}(t) = 2t/(1 + t)$ (purple dotted line). Any MC function $f(t)$ satisfies ${f_{min}}(t) \leq f(t) \leq {f_{max}}(t)$, i.e., its graph falls into the shaded area, as shown in the particular case of the MC function ${f_{WY}}(t) = (1/4){(\sqrt{t} + 1)^2}$ corresponding to the Wigner-Yanase information metric (blue dashed line). All the MC functions collapse to $1$ at $t=1$. This reflects the fact that the corresponding Riemannian metrics are regular on the maximally mixed state.}
\label{FIG001}
\end{figure}

As pointed out by Kubo and Ando~\cite{Kubo_Ando_Math_Ann_03_246}, among the MC functions there exists a minimal one, ${f_{min}}(t) = 2t/(1 + t)$, and a maximal one, ${f_{max}}(t) = (1 + t)/2$, such that a generic MC function $f(t)$ must satisfy ${f_{min}}(t)\leq f(t) \leq {f_{max}}(t)$. Interestingly, the maximal MC function is the one corresponding to the celebrated quantum Fisher information metric, whereas the Wigner-Yanase information metric corresponds to an intermediate MC function, ${f_{WY}}(t) = (1/4){(\sqrt{t} + 1)^2}$, as illustrated in Fig.~\ref{FIG001}.

Each of these metrics plays a fundamental role in quantum information theory since the corresponding geodesic length $\mathcal{L}$, being by construction contractive under quantum stochastic maps, represents a bona fide measure of distinguishability over the quantum state space. However, finding such geodesic distance is unfortunately a very hard task in general. In fact, analytic expressions are known only for the geodesic distance related to the quantum Fisher information metric~\cite{1993_RepMathPhys_33_253,*1995_RepMathPhys_36_461},
\begin{equation}\label{eq:BuresAngle}
\mathcal{L}^{QF}(\rho,\sigma) = \arccos\left[\sqrt{{F}({\rho},{\sigma})} \, \right] ~,
\end{equation}
where ${F}({\rho},{\sigma}) = \left(\text{Tr}\left[\sqrt{\sqrt{\rho}\sigma\sqrt{\rho}} \, \right]\right)^2$ is the Uhlmann fidelity, and for the one related to the Wigner-Yanase information metric~\cite{2003_JMathPhys_44_3752},
\begin{equation}\label{eq:HellingerAngle}
\mathcal{L}^{WY}(\rho,\sigma) = \arccos\, [{A}(\rho,\sigma)] ~,
\end{equation}
where ${A}(\rho,\sigma) = \text{Tr} \left(\sqrt{\rho}\sqrt{\sigma}\right)$ is known as quantum affinity.

\section{Generalized Geometric Quantum speed limits}
\label{sec:sectionIII}

We are now ready to present our main result, that is, a family of geometric QSLs which hold for any physical process and are in one to one correspondence with the contractive Riemannian metrics defined on the set of quantum states. The most general dynamical evolution of an initial state $\rho_0$ can be written in the Kraus decomposition as ${\rho_{\lambda}} = {\mathcal{E}_{\lambda}}[\, {\rho_0}]={\sum_j}{K_j^{\lambda}} \, {\rho_0}{K_j^{\lambda\dagger}}$, where $\{{K_j^{\lambda}}\}$ are operators satisfying ${\sum_j}{K_j^{\lambda\dagger}}{K_j^{\lambda}} = \mathbb{I}$ and depending on a set $\lambda = \{{\lambda_1},{\lambda_2},\ldots,{\lambda_r}\}$ of $r$ parameters which are encoded into the input state $\rho_0$, in such a way that $\rho_\lambda$ depends analytically on each parameter $\lambda_{\mu}$ ($\mu = 1,\ldots,r$). In the unitary case, the evolution is given in particular by ${\mathcal{E}_{\lambda}}[\, {\rho_0}] = {U_{\lambda}} \, {\rho_0} {U_{\lambda}^{\dagger}}$, where ${U_{\lambda}}$ is a multiparameter family of unitary operators, fulfilling ${U_{\lambda}}{U_{\lambda}^{\dagger}} ={U_{\lambda}^{\dagger}}{U_{\lambda}}= \mathbb{I}$. In this case, the observables ${H^{\lambda}_{\mu}} = {-i\hbar}{U_{\lambda}} \, {\partial_{\mu}}{U_{\lambda}^{\dagger}}$, with ${\partial_{\mu}} \equiv {\partial}/\partial{\lambda_{\mu}}$, are the generators driving the dynamics.

Consider a dynamical evolution $\rho_{\lambda}$ in which the set of parameters $\lambda$ is changed analytically from the initial configuration $\lambda_I$ to the final one $\lambda_F$. Geometrically, this evolution draws a path $\gamma$ in the quantum state space connecting $\rho_{\lambda_I}$ and $\rho_{\lambda_F}$ whose length is given by the line integral ${\ell^f_{\gamma}} = {\int_{\gamma}}\, ds$ and depends on the chosen metric $\textbf{g}^f$ (see Fig.~\ref{figQS}). Since $\gamma$ is an arbitrary path between $\rho_{\lambda_I}$ and $\rho_{\lambda_F}$, its length need not be the shortest one, which is instead given by the geodesic length $\mathcal{L}^f(\, {\rho_{\lambda_I}},{\rho_{\lambda_F}})$ between $\rho_{\lambda_I}$ and $\rho_{\lambda_F}$. Therefore the latter represents a lower bound for the length of the path drawn by the above dynamical evolution. This observation will play a crucial role in the imminent derivation of our family of QSLs, in analogy with Refs.~\cite{1990_PhysRevLett_65_1697} and~\cite{2013_PhysRevLett_110_050402}.

Since the density operator $\rho_{\lambda}$ evolves analytically with respect to the parameters $\lambda$, we can write
\begin{equation}
 \label{eq:mrzcncpetz0002}
d{\rho_{\lambda}} = {\sum_{\mu = 1}^r} \, {\partial_{\mu}}{\rho_{\lambda}} \, d{\lambda_{\mu}}.
\end{equation}
Let ${\rho_{\lambda}} = {\sum_j}{p_j}|{j}\rangle\langle{j}|$ be the spectral decomposition of $\rho_\lambda$, with $0 < {p_j} \leq 1$ and ${\sum_j}{p_j} = 1$. We note that both the eigenvalues $p_j$ and eigenstates $|j\rangle$ of $\rho_\lambda$ may depend on the set of parameters $\lambda$, i.e. ${p_j} \equiv {p_j}(\lambda)$ and $|j\rangle \equiv |j(\lambda)\rangle$, so that
\begin{equation}
 \label{eq:mrzcncpetz000290a}
{\partial_{\mu}}{\rho_{\lambda}} = {\sum_{j}}\{({\partial_{\mu}}{p_j})|j\rangle\langle{j}\, | +
{p_j}[({\partial_{\mu}}|j\rangle)\langle{j}\, | + |j\rangle({\partial_{\mu}}\langle{j}\, |)]\} ~,
\end{equation}
and thus
\begin{equation}
 \label{eq:mrzcncpetz00029}
\langle{j}|{\partial_{\mu}}{\rho_{\lambda}}|{l}\rangle = {\delta_{jl}}{\partial_{\mu}}{p_j} + ({p_l} - {p_j})\langle{j}|{\partial_{\mu}}|{l}\rangle ~,
\end{equation}
where we used the identity $({\partial_{\mu}}\langle{j}|)|l\rangle = -\langle{j}|{\partial_{\mu}}|{l}\rangle$. Combining Eq.~\eqref{eq:mrzcncpetz0002} and Eq.~\eqref{eq:mrzcncpetz00029}, we get
\begin{equation}
 \label{eq:mrzcncpetz0005}
\langle{j}|d{\rho_{\lambda}}|l\rangle = {\sum_{\mu = 1}^r} \, [{\delta_{jl}}{\partial_{\mu}}{p_j} + i({p_j} - {p_l}){\mathcal{A}_{jl}^{\mu}}]d{\lambda_{\mu}}  ~,
\end{equation}
where we define ${\mathcal{A}_{jl}^{\mu}} \equiv i\langle{j}|{\partial_{\mu}}|{l}\rangle$. By using  Eq.~\eqref{eq:mrzcncpetz0005}, in the case of $j = l$ we get
\begin{equation}
 \label{eq:mrzcncpetz0006}
{|\langle{j}|d{\rho_{\lambda}}|j\rangle|^2} = {\sum_{\mu,\nu = 1}^r} \, {\partial_{\mu}}{p_j}{\partial_{\nu}}{p_j} \,
d{\lambda_{\mu}}d{\lambda_{\nu}},
\end{equation}
whereas in the case of $j \neq l$, by using the fact that $d{\rho_{\lambda}}$ is hermitian, we obtain
\begin{equation}
 \label{eq:mrzcncpetz0007}
{|\langle{j}|d{\rho_{\lambda}}|l\rangle|^2} = {\sum_{\mu,\nu = 1}^r} \, {({p_j} - {p_l})^2}{\mathcal{A}_{jl}^{\mu}}{\mathcal{A}_{lj}^{\nu}} \,
d{\lambda_{\mu}}d{\lambda_{\nu}} ~.
\end{equation}
Finally, by substituting Eqs.~\eqref{eq:mrzcncpetz0006} and~\eqref{eq:mrzcncpetz0007} into Eq.~\eqref{eq:mrzcncpetz0001}, the squared infinitesimal length $ds^2$ between $\rho_\lambda$ and $\rho_\lambda + d\rho_\lambda$ according to any contractive Riemannian metric $\textbf{g}^f$ becomes
\begin{equation}
 \label{eq:mrzcncpetz0008}
{ds^2} = {\sum_{\mu,\nu=1}^r} \, {g^f_{\mu\nu}}d{\lambda_{\mu}}d{\lambda_{\nu}} ~,
\end{equation}
where
\begin{equation}
 \label{eq:mrzcncpetz000901}
{g^f_{\mu\nu}} = {\mathcal{F}_{\mu\nu}} + {\mathcal{Q}^f_{\mu\nu}} ~,
\end{equation}
with
\begin{equation}
 \label{eq:mrzcncpetz00090101}
{\mathcal{F}_{\mu\nu}} =  \frac{1}{4}{\sum_{j}}\frac{{\partial_{\mu}}{p_j}
{\partial_{\nu}}{p_j}}{p_j} ~,
\end{equation}
and
\begin{equation}
 \label{eq:mrzcncpetz00090102}
{\mathcal{Q}^f_{\mu\nu}} =  \frac{1}{2}{\sum_{j < \, l}} \, {c^f} ({p_j},{p_l})
{({p_j} - {p_l})^2}{\mathcal{A}_{jl}^{\mu}}{\mathcal{A}_{lj}^{\nu}} ~,
\end{equation}
referring to, respectively, the contribution of the populations of $\rho_\lambda$ and of the coherences of $d\rho_\lambda$ to the contractive Riemannian metric tensor $g^f_{\mu\nu}$. Herein we restrict to the case where the parameters $\lambda$ are time-dependent, ${\lambda_{\mu}} = {\lambda_{\mu}}\, (t)$, for $\mu = 1,\ldots,r$, and choose the parametrization $t \in [0,\tau] \rightarrow {\lambda}\, (t)$ such that ${\lambda_I} = \lambda\, (0)$ and ${\lambda_F} = \lambda\, (\tau)$, where $\tau$ is the evolution time. Now, being the geodesic distance ${\mathcal{L}^f}({\rho_{0}},{\rho_{\tau}})$ between the initial and final state, $\rho_{0}$ and $\rho_{\tau}$, a lower bound to the length $\ell^f_\gamma({\rho_{0}},{\rho_{\tau}})= \int_\gamma ds = {\int_0^{\tau}}\, dt \, (ds/dt)$ of the path $\gamma$ followed by the evolved state $\rho_t$ when going from $\rho_{0}$ to $\rho_{\tau}$, we have
\begin{equation}
\label{eq:moroimpr003}
{\mathcal{L}^f}({\rho_{0}},{\rho_{\tau}}) \leq {{\ell}^f_{\gamma}}({\rho_0},{\rho_{\tau}}) ~,
\end{equation}
where
\begin{equation}
\label{eq:moroimpr004}
{{\ell}^f_{\gamma}}({\rho_0},{\rho_{\tau}}) = {\int_0^{\tau}}dt \, \sqrt{{\sum_{\mu,\nu = 1}^r}\,{g^f_{\mu\nu}}\frac{d{\lambda_{\mu}}}{dt}\frac{d{\lambda_{\nu}}}{dt}} ~.
\end{equation}

Equation~\eqref{eq:moroimpr003} represents the anticipated infinite family of generalized geometric QSLs and is the central result of this paper. Any possible contractive Riemannian metric $\textbf{g}^f$ on the quantum state space, and so any possible bona fide geometric quantifier of distinguishability between quantum states, gives rise to a different QSL. More precisely, we have that both the geodesic distance appearing in the left hand side and the quantity ${{\ell}^f_{\gamma}}({\rho_0},{\rho_{\tau}})$ being in the right hand side of Eq.~\eqref{eq:moroimpr003} depend on the chosen contractive Riemannian metric, specified by a MC function $f$. In particular, by restricting to the celebrated quantum Fisher information metric, we recover the  QSL introduced in Ref.~\cite{2013_PhysRevLett_110_050402}.

It is intuitively clear that the contractive Riemannian metric whose geodesic is most tailored to the given dynamical evolution is the one that gives rise to the tightest lower bound to the evolution time as expressed in Eq.~\eqref{eq:moroimpr003}. In order to determine how much a certain geometric QSL is saturated, i.e.~its {\it tightness}, we will consider the relative difference
\begin{equation}\label{eq:delta}
\delta^f_\gamma := \frac{{{\ell}^f_{\gamma}}({\rho_0},{\rho_{\tau}})-{\mathcal{L}^f}({\rho_{0}},{\rho_{\tau}})}{{\mathcal{L}^f}({\rho_{0}},{\rho_{\tau}})},
\end{equation}
that quantifies how much the dynamical evolution $\gamma$ differs from a geodesic with respect to the considered metric $\textbf{g}^f$.

By  minimizing the quantity $\delta^f_\gamma$ over all contractive Riemannian metrics, i.e., over all MC functions $f$, one has a criterion to identify in principle the tightest geometric QSL, of the form given in Eq.~(\ref{eq:moroimpr003}),  for any given dynamics $\gamma$. Formally, labelling by $f^{\star}_\gamma$ the optimal metric for the dynamics $\gamma$, the tightest possible geometric QSL is therefore defined by
\begin{equation}
\label{eq:optimal}
{\mathcal{L}^{f^{\star}_\gamma}}({\rho_{0}},{\rho_{\tau}}) \leq {{\ell}^{f^{\star}_\gamma}_{\gamma}}({\rho_0},{\rho_{\tau}}) ~, \mbox{\ \ with \ \    $f^{\star}_\gamma$\ \  such that\ \ } \inf_{f} \delta^f_\gamma \equiv \delta^{f^{\star}_\gamma}_\gamma\,,
\end{equation}
where the minimization is over all MC functions $f$.

Finding this minimum is, however, a formidable problem, which is made all the more difficult by the fact that the quantum Fisher information metric and the Wigner-Yanase information metric are the only contractive Riemannian metrics whose geodesic lengths are analytically known for general dynamics (as previously remarked).

Nevertheless, in this paper we will move the first steps forward towards addressing such general problem, by restricting the optimization in Eq.~(\ref{eq:optimal}) primarily over these two paradigmatic and physically significant examples of contractive Riemannian metrics, namely the quantum Fisher information and the Wigner-Yanase skew information. Quite remarkably, this restriction will be enough to reveal how the choice of the quantum Fisher information metric, though ubiquitous in the existing literature, is only a special case which does not always provide the tightest lower bound. On the contrary, we will show how the  Wigner-Yanase skew information metric can systematically produce tighter bounds in a number of situations of practical relevance for quantum information and quantum technologies, in particular in open system evolutions.

\section{Examples}
\label{sec:examplesPD_AD}

In this section we will apply our general formalism to present and analyze QSLs based primarily on the quantum Fisher information and the Wigner-Yanase skew information in a selection of unitary and nonunitary physical processes. This will serve the purpose to illustrate how the choice of a particular bona fide geometric measure of distinguishability on the quantum state space affects the QSLs, therefore providing a guidance to exploit the freedom in this choice to obtain the tightest bounds in practical scenarios.

\subsection{Unitary dynamics}

We start by restricting ourselves to a closed quantum system, so that our initial state $\rho_0$ undergoes a unitary evolution ${\rho_{\lambda}} = {U_{\lambda}} \, {\rho_0} {U_{\lambda}^{\dagger}}$. Since the eigenvalues $p_j$ of a unitarily evolving state are constant, ${\partial_{\mu}}{p_j} = 0$,  we have that ${\mathcal{F}_{\mu\nu}} = 0$, and thus ${g^f_{\mu\nu}} = {\mathcal{Q}^f_{\mu\nu}}$, along the curve $\gamma$ drawn by the evolved state $\rho_\lambda$. In other words, the coherences of $d\rho_\lambda$ drive the evolution of a closed quantum system. Moreover, one can easily see that $\mathcal{A}_{jl}^{\mu} = (1/\hbar)\langle{j}|\Delta{H^{\lambda}_{\mu}}|{l}\rangle$, where $\Delta{H^{\lambda}_{\mu}} = {H^{\lambda}_{\mu}} - \langle{H^{\lambda}_{\mu}}\rangle$, $\langle{H^{\lambda}_{\mu}}\rangle=\mbox{Tr}(\rho_\lambda H^{\lambda}_\mu)$, and ${H^{\lambda}_{\mu}} = {-i\hbar}{U_{\lambda}} \, {\partial_{\mu}}{U_{\lambda}^{\dagger}}$ are the generators of the dynamics, so that
\begin{equation}
 \label{eq:qcdynqsl001}
{g^f_{\mu\nu}} = \frac{1}{2{\hbar^2}}{\sum_{j < \, l}} \, {c^f}({p_j},{p_l}){({p_j} - {p_l})^2}{\langle{j}|\Delta{H^{\lambda}_{\mu}}|{l}\rangle\langle{l}|\Delta{H^{\lambda}_{\nu}}|{j}\rangle} ~.
\end{equation}
In the following subsections we will focus on, respectively, the quantum Fisher information metric and the Wigner-Yanase information metric.

\subsubsection{Quantum Fisher information metric}

The quantum Fisher information metric corresponds to the MC function $f(t) = (1 + t)/2$, so that ${c^f}(x,y) = 2/(x + y)$ and  Eq.~\eqref{eq:qcdynqsl001} becomes
\begin{equation}
 \label{eq:qcdynqsl002}
{g^{QF}_{\mu\nu}} = \frac{1}{2{\hbar^2}}{\sum_{j,l}} \, \frac{({p_j} - {p_l})^2}{{p_j} + {p_l}}{\langle{j}|\Delta{H^{\lambda}_{\mu}}|{l}\rangle\langle{l}|\Delta{H^{\lambda}_{\nu}}|{j}\rangle} ~.
\end{equation}
Moreover, by using the following straightforward inequality
\begin{equation}
 \label{eq:mrzcncpetz00012}
\frac{({p_j} - {p_l})^2}{{p_j} + {p_l}} \leq {p_j} + {p_l} ~,
\end{equation}
we get
\begin{align}
 \label{eq:qcdynqsl003}
{g^{QF}_{\mu\nu}} &\leq \frac{1}{2{\hbar^2}}{\sum_{j, l}} \, {({p_j} + {p_l})}{\langle{j}|\Delta{H^{\lambda}_{\mu}}|{l}\rangle\langle{l}|\Delta{H^{\lambda}_{\nu}}|{j}\rangle} \nonumber \\
&= {\hbar^{-2}}\mathscr{C}(\Delta{H^{\lambda}_{\mu}},\Delta{H^{\lambda}_{\nu}}) ~,
\end{align}
where $\mathscr{C}(\Delta{H^{\lambda}_{\mu}},\Delta{H^{\lambda}_{\nu}}) := ({1}/{2})\text{Tr}[\rho_\lambda\{\Delta{H^{\lambda}_{\mu}},\Delta{H^{\lambda}_{\nu}}\}]$ is the symmetrized covariance of $\Delta H^{\lambda}_{\mu}$ and $\Delta H^{\lambda}_{\nu}$ with respect to the evolved state, which reduces to the squared variance of the operator ${H^{\lambda}_{\mu}}$ when $\mu = \nu$, i.e. $\mathscr{C}(\Delta{H^{\lambda}_{\mu}},\Delta{H^{\lambda}_{\mu}}) = \text{Tr}[\rho_\lambda{(\Delta{H^{\lambda}_{\mu}})^2}] = \langle{({H^{\lambda}_{\mu}})^2}\rangle - {\langle{H^{\lambda}_{\mu}}\rangle^2}$. By substituting the inequality \eqref{eq:qcdynqsl003} into Eq.~\eqref{eq:moroimpr003} we get the new bound
\begin{eqnarray}
 \label{eq:qcdynqsl00302}
&&{\mathcal{L}^{QF}}({\rho_0},{\rho_{\tau}}) \nonumber \\
&&\leq \frac{1}{\hbar}{\int_0^{\tau}}dt \, \sqrt{{\sum_{\mu,\nu = 1}^r}\,\mathscr{C}(\Delta{H^{\lambda}_{\mu}},\Delta{H^{\lambda}_{\nu}})\frac{d{\lambda_{\mu}}}{dt}\frac{d{\lambda_{\nu}}}{dt}} ~.
\end{eqnarray}

Although the QSL in Eq.~\eqref{eq:qcdynqsl00302} applies to the very general  $r$-parameter case, let us restrict for simplicity to the one-parameter case where $\lambda = t$. Consequently, we have that ${H_{\mu}^{\lambda}} \rightarrow {H_t} = {-i\hbar}{U_t} \, {\partial_t}{U_t^{\dagger}}$ and that the symmetrized covariance just reduces to the variance of the observable ${H_t}$ generating the dynamics of the system. Therefore, Eq. \eqref{eq:qcdynqsl00302} turns into the simpler bound
\begin{align}
 \label{eq:qcdynqsl00303}
\tau^{-1}{\mathcal{L}^{QF}}({\rho_0},{\rho_{\tau}}) &\leq \frac{1}{\hbar\,\tau}{\int_0^{\tau}}dt\, \sqrt{\langle{{H_t^2}}\rangle - {\langle{H_t}\rangle^2}} \nonumber \\
&= {\hbar^{-1}}{\overline{\Delta{E}}} ~,
\end{align}
where $\overline{\Delta{E}} := \tau^{-1}{\int_0^{\tau}}dt\, \sqrt{\langle{{H_t^2}}\rangle - {\langle{H_t}\rangle^2}}$ is the mean variance of the generator $H_t$. The following QSL is thus obtained
\begin{equation}
 \label{eq:qcdynqsl00304}
\tau \geq \frac{\hbar}{\overline{\Delta{E}}}{\mathcal{L}^{QF}}({\rho_0},{\rho_{\tau}}) ~.
\end{equation}
It is worth emphasizing that the bound in Eq.~\eqref{eq:qcdynqsl00304} applies to arbitrary initial and final mixed states and generic time-dependent generators of the dynamics. Moreover, we can immediately see that it exactly coincides with the one reported in Ref.~\cite{uhlmann1992energy} and reduces to a MT-like bound, when further restricting to the case of a time-independent generator of the dynamics.

\subsubsection{Wigner-Yanase information metric}

The Wigner-Yanase information metric corresponds to the MC function $f(t) = (1/4){(\sqrt{t} + 1)^2}$, so that ${c^f}(x,y) = 4/{(\sqrt{x} + \sqrt{y})^2}$ and Eq.~\eqref{eq:qcdynqsl001} becomes
\begin{align}
 \label{eq:qcdynqsl004}
{g^{WY}_{\mu\nu}} &= \frac{2}{\hbar^2}{\sum_{j < l}} \, {\left(\frac{{p_j} - {p_l}}{\sqrt{p_j} + \sqrt{p_l}}\right)^2}{\langle{j}|\Delta{H^{\lambda}_{\mu}}|{l}\rangle\langle{l}|\Delta{H^{\lambda}_{\nu}}|{j}\rangle} \nonumber \\
&= \frac{1}{\hbar^2}{\sum_{j, l}} \, {(\sqrt{p_j} - \sqrt{p_l})^2}{\langle{j}|\Delta{H^{\lambda}_{\mu}}|{l}\rangle\langle{l}|\Delta{H^{\lambda}_{\nu}}|{j}\rangle} \nonumber \\
&= - \frac{1}{\hbar^2}\text{Tr}({[\sqrt{\rho},\Delta{H^{\lambda}_{\mu}}]}{[\sqrt{\rho},\Delta{H^{\lambda}_{\nu}}]}) \nonumber \\
&= \frac{2}{\hbar^2}\mathcal{C}\, (\Delta{H^{\lambda}_{\mu}},\Delta{H^{\lambda}_{\nu}}) ~,
\end{align}
where $\mathcal{C}(\Delta{H^{\lambda}_{\mu}},\Delta{H^{\lambda}_{\nu}}) := -(1/2)\text{Tr}({[\sqrt{\rho_\lambda},\Delta{H^{\lambda}_{\mu}}]}{[\sqrt{\rho_\lambda},\Delta{H^{\lambda}_{\nu}}]})$ reduces to the skew information between the evolved state and $\Delta{H^{\lambda}_{\mu}}$ when $\mu=\nu$ , $\mathcal{C}(\Delta{H^{\lambda}_{\mu}},\Delta{H^{\lambda}_{\mu}}) = \mathcal{I}(\rho_\lambda,\Delta{H^{\lambda}_{\mu}}) :=  -(1/2)\text{Tr}({[\sqrt{\rho_\lambda},\Delta{H^{\lambda}_{\mu}}]}^2)$~\cite{Gibilisco_Isola_AnnInstStatMath_147_59}. By putting Eq.~\eqref{eq:qcdynqsl004} into the bound in Eq.~\eqref{eq:moroimpr003},  we get
\begin{eqnarray}
 \label{eq:qcdynqsl006}
&&{\mathcal{L}^{WY}}({\rho_0},{\rho_{\tau}}) \nonumber \\
&&\leq \frac{\sqrt{2}}{\hbar}{\int_0^{\tau}}dt \, \sqrt{{\sum_{\mu,\nu = 1}^r}\,\mathcal{C}\, (\Delta{H^{\lambda}_{\mu}},
\Delta{H^{\lambda}_{\nu}})\frac{d{\lambda_{\mu}}}{dt}\frac{d{\lambda_{\nu}}}{dt}} ~.
\end{eqnarray}

For simplicity, let us again analyze the one-parameter case, where $\lambda = t$, ${H_{\mu}^{\lambda}} \rightarrow {H_t} = {-i\hbar}{U_t} \, {\partial_t}{U_t^{\dagger}}$ and $\mathcal{C}$ reduces to the skew information $\mathcal{I}(\, {\rho_t},{H_t})$ between the evolved state ${\rho_t}$ and the observable ${H_t}$ generating the dynamics of the system. Therefore, the bound in Eq.~\eqref{eq:qcdynqsl006} turns into
\begin{align}
 \label{eq:qcdynqsl00601}
\tau^{-1}{\mathcal{L}^{WY}}({\rho_0},{\rho_{\tau}}) &\leq \frac{\sqrt{2}}{\hbar}\frac{1}{\tau}{\int_0^{\tau}}dt\, \sqrt{\mathcal{I}(\, {\rho_t},{H_t})} \nonumber \\
&= \frac{\sqrt{2}}{\hbar}{\overline{\sqrt{\mathcal{I}}}} ~,
\end{align}
where we define $\overline{\sqrt{\mathcal{I}}} := \tau^{-1}{\int_0^{\tau}}dt\, \sqrt{\mathcal{I}(\, {\rho_t},{H_t})}$ as the mean skew information between the evolved state and the generator of the evolution. The QSL thus becomes
\begin{equation}
 \label{eq:qcdynqsl00602}
\tau \geq \frac{\hbar}{\sqrt{2}\overline{\sqrt{\mathcal{I}}}}{\mathcal{L}^{WY}}({\rho_0},{\rho_{\tau}}) ~.
\end{equation}
As reported by Luo~\cite{PhysRevA.73.022324}, the skew information $\mathcal{I}(\, {\rho_t},{H_t})$ is upper bounded by the variance of the observable $H_t$, $\mathcal{I}(\, {\rho_t},{H_t}) \leq \langle{{H_t^2}}\rangle - {\langle{H_t}\rangle^2}$, so that $\overline{\sqrt{\mathcal{I}}} \leq \overline{\Delta{E}}$ and
\begin{equation}
 \label{eq:qcdynqsl00603}
\tau \geq \frac{\hbar}{\sqrt{2}\overline{\sqrt{\mathcal{I}}}}{\mathcal{L}^{WY}}({\rho_0},{\rho_{\tau}}) \geq
\frac{1}{\sqrt{2}}\frac{\hbar}{\overline{\Delta{E}}}{\mathcal{L}^{WY}}({\rho_0},{\rho_{\tau}}) ~.
\end{equation}
The latter QSL strongly resembles the bound expressed in Eq.~\eqref{eq:qcdynqsl00304} and emerging from the quantum Fisher information metric, with the difference lying in the fact that we are now adopting the Hellinger angle instead of the Bures angle and a $\sqrt{2}$ factor appears in the denominator. However, when the initial and final states commute, we have that the corresponding fidelity and affinity coincide, $F({\rho_0},{\rho_{\tau}}) = A({\rho_0},{\rho_{\tau}})$, and so the Bures angle is equal to the Hellinger angle, which implies that in this case the bound emerging from the Wigner-Yanase metric is less tight than the one corresponding to the quantum Fisher information by a factor of $1/\sqrt{2}$. 

The above result could be intuitively expected due to the strict hierarchy respected by the MC functions corresponding to the two adopted metrics. To put such an intuition on rigorous grounds, in Appendix~\ref{sec:U} we prove that the geometric QSL corresponding to the quantum Fisher information metric, as expressed directly by Eq.~\eqref{eq:moroimpr003}, is indeed tighter than the one corresponding to the Wigner-Yanase information metric, when considering {\it any} single-qubit unitary dynamics. However, we leave it as an open question to assess whether this is still the case when considering higher dimensional quantum systems, or other contractive Riemannian metrics in place of the Wigner-Yanase one.

Quite surprisingly, we will show instead in the next section that, for the realistic and more general case of nonunitary dynamics, the hierarchy of the MC functions does not automatically translate anymore into a hierarchy of tightness for the corresponding QSLs, not even in the case of a single qubit. This will reveal original consequences of our analysis in practically relevant scenarios.

\subsection{Nonunitary dynamics}

We will now consider two paradigmatic examples of nonunitary physical processes acting on a single qubit: dephasing and amplitude damping.

\subsubsection{Parallel and transversal dephasing channels}
\label{sec:sectionphasenew}

\begin{figure*}[t]
\centering
\includegraphics[scale=0.43]{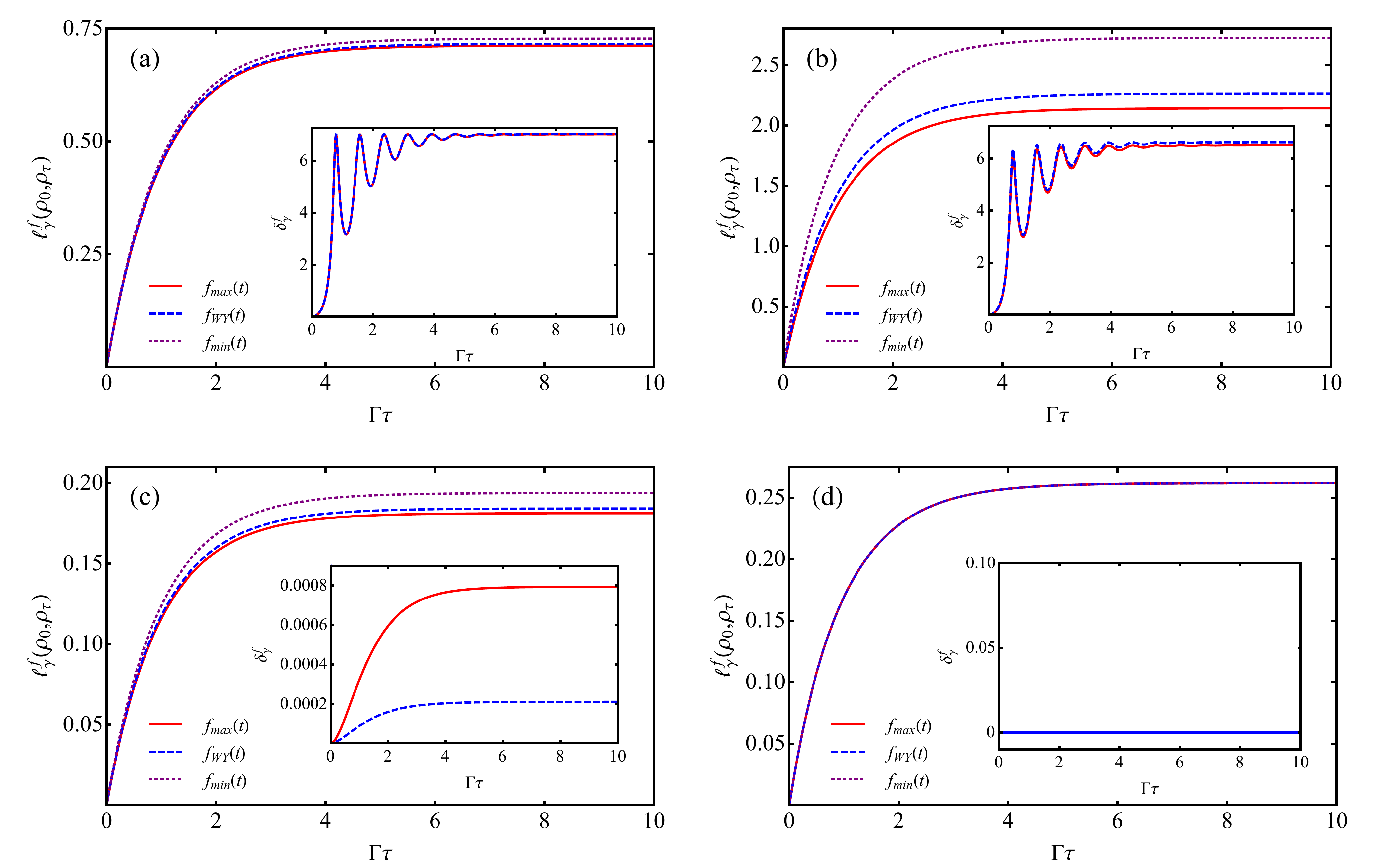}
\caption{(Color online). Evolution path lengths $\ell^f_\gamma$ for parallel dephasing processes by considering the contractive Riemannian metrics corresponding to the following MC functions: ${f_{QF}}(t)={f_{max}}(t)$ (red solid line), ${f_{WY}}(t)$ (blue dashed line) and ${f_{min}}(t)$ (purple dotted line) for (\textbf{a}) ${r_0} = 1/4$, $\theta_0 = \pi/4$, $\beta = 8$ (\textbf{b}) ${r_0} = 3/4$, $\theta_0 = \pi/4$,  $\beta = 8$ (\textbf{c}) ${r_0} = 1/2$, $\theta_0 = \pi/4$,  $\beta = 0$ and (\textbf{d}) ${r_0} = 1/2$, $\theta_0 = \pi/2$,  $\beta = 0$. The insets in each panel show the relative difference ${\delta^f_{\gamma}}$, Eq.~(\ref{eq:delta}), for parallel dephasing processes by considering the quantum Fisher information metric (red solid line) and the Wigner-Yanase information metric (blue dashed line); such a relative difference can be regarded as an indicator of the tightness of the bounds (the smaller $\delta^f_\gamma$, the tighter the bounds).}
\label{FIG005}
\end{figure*}
Let us denote by ${\rho} = (1/2)(\mathbb{I} + {\vec{r}}\cdot\vec{\sigma})$ the Bloch sphere representation of an arbitrary single qubit state $\rho$, where ${\vec{r}} = \{r\sin\theta\cos\phi,r\sin\theta\sin\phi,r\cos\theta\}$ is the Bloch vector, with $r\in[0,1]$, $\theta\in[0,\pi]$ and $\phi\in[0,2\pi[$, while $\mathbb{I}$ denotes the $2\times 2$ identity matrix and $\vec{\sigma} = \{{\sigma_1},{\sigma_2},{\sigma_3}\}$ is the vector of Pauli matrices.

Let us now consider a noisy evolution of this state governed by a master equation of Lindblad form
\begin{equation}
\frac{\partial \rho(t)}{\partial t} = \mathscr{H}(\rho) + \mathscr{L}(\rho) ~,
\end{equation}
where $\mathscr{H}(\rho)= -i[H,\rho]$ describes the unitary evolution governed by a Hamiltonian $H$ while $\mathscr{L}(\rho)$ is the Liouvillian that describes the noise. We further consider as Hamiltonian $H=\frac{\omega_0}{2} \sigma_3$, where $\omega_0$ is the unitary frequency, and as Liouvillian
\begin{equation}
\mathscr{L}(\rho) = - \frac{\Gamma}{2}\left(\rho - \sum_{i=1}^3 \alpha_i \sigma_i \rho \sigma_i\right) ~,
\end{equation}
where $\Gamma$ is the decoherence rate and $\alpha_i\geq 0$ with $\sum_i\alpha_i=1$.

We can identify two main modalities of dephasing noise. When $\alpha_3=1$, the dephasing happens in the same basis as the one specifying the Hamiltonian of our system, a case that can be referred to as `parallel dephasing'. When instead $\alpha_1=1$, the dephasing occurs in a basis orthogonal to the one of the Hamiltonian, leading to the situation typically referred to as `transversal dephasing'~\cite{Chaves2013Noisy,Chaves2015PRX}. We will explore these two cases separately.

\paragraph{Parallel dephasing.}
The parallel dephasing noise lets an initial state $\rho_0$ evolve as $\rho_{t}={\sum_{j=0}^1}{K_j}\rho_{0}{K_j^{\dagger}}$, where
\begin{equation}
 \label{eq:newsplkraus0001}
{K_0} =  \sqrt{{q_+}}\left(\begin{matrix} {e^{-i{\omega_0}t/2}} & 0 \\ 0 & {e^{i{\omega_0}t/2}} \end{matrix}\right) ~,~
{K_1} =  \sqrt{{q_-}}\left(\begin{matrix} {e^{-i{\omega_0}t/2}} & 0 \\ 0 & -{e^{i{\omega_0}t/2}} \end{matrix}\right)
\end{equation}
are the Kraus operators, and ${q_{\pm}} = (1 \pm {q_t})/2$ with ${q_t} = {e^{-\Gamma t}}$~\cite{Nielsen_Chuang_infor_geom}. Notice that the Kraus operators satisfy not only ${\sum_{j=0}^1}{K_j^{\dagger}}{K_j} = \mathbb{I}$ but also ${\sum_{j=0}^1}{K_j}{K_j^{\dagger}} = \mathbb{I}$, as such a channel is unital. The effect of parallel dephasing is exactly the same as the one of phase flip and consists in shrinking the Bloch sphere onto the $z$-axis of states diagonal in the computational basis, which are instead left invariant. Moreover, $\omega_0$ describes the rotation frequency around the $z$-axis. One can easily see that the evolved state $\rho_{t}$ has the following spectral decomposition
\begin{equation}
\label{eq:newspectralphasedamp001}
{\rho_{t}} = {\sum_{j=\pm}}\, {p_j}{|\theta_{t},\phi_{t}\rangle_j}{\langle\theta_{t},\phi_{t}|_j} ~,
\end{equation}
where ${p_{\pm}} = ({1}/{2})(1 \pm {r_0}\, \xi_{t})$  and
\begin{equation}
\label{eq:neweigsphasedamp002}
{|{\theta_t},{\phi_t}\rangle_{\pm}} = \frac{1}{N_{\pm}}\left[(\cos{\theta_0} \pm {\xi_t})|0\rangle + {e^{i({\omega_0}t + {\phi_0})}}{q_t}\sin{\theta_0}|1\rangle\right] ~,
\end{equation}
with ${\xi_t} = \sqrt{{\cos^2}{\theta_0} + {q_t^2}{\sin^2}{\theta_0}}$ and $N_{\pm}$ a normalization constant. By putting the above equations into Eqs.~\eqref{eq:mrzcncpetz00090101} and~\eqref{eq:mrzcncpetz00090102} one obtains, respectively,
\begin{equation}
 \label{eq:newfisherphasedamp001}
{\mathcal{F}} = \frac{{r_0^2}{q_t^2}{\sin^4}{\theta_0}\, {({d{q_t}}/{dt})^2}}{4 \xi_t^2\, (1 - {r_0^2}\xi_t^2)}
\end{equation}
and
\begin{equation}
 \label{eq:newcohphasedamp001}
{\mathcal{Q}^f} = \frac{1}{8}\left[{\omega_0^2}{q_t^2} + \frac{{\cos^2}{\theta_0}\, {({d{q_t}}/{dt})^2}}{\xi_t^2}\right]{r_0^2}{\sin^2}{\theta_0}\, {c^f}({p_+},{p_-}) ~.
\end{equation}

The contractive Riemannian metric $g^f=\mathcal{F}+\mathcal{Q}^f$ can be interpreted as the speed of evolution of $\rho_{t}$. Equation~\eqref{eq:newfisherphasedamp001}, which corresponds to the contribution to $g^f$ common to all the MC family, is identically zero for all the initial states such that $\theta_0$ is either $0$ or $\pi$, that are all the incoherent states lying on the $z$-axis of the Bloch sphere (with density matrices diagonal in the computational basis), which are indeed left unaffected by the parallel dephasing dynamics. Although $\mathcal{F}$ is a function of the initial purity $r_0$ and of time, it does not depend on the initial azimuthal angle $\phi_0$ since the eigenvalues of the evolved state $p_j$ do not depend on $\phi_0$. Equation~\eqref{eq:newcohphasedamp001}, which instead describes the truly quantum contribution to the speed of evolution $g^f$ and depends on the specific choice of the MC function $f$, is identically zero for all the incoherent initial states such that $\theta_0$ is either $0$ or $\pi$. Notice that in the case ${\theta_0} = \pi/2$, for initial states lying in the equatorial $xy$-plane, $\mathcal{Q}^f$ is nonzero only when the frequency $\omega_0$ is also nonzero. Interestingly,  $\mathcal{Q}^f$ does not depend on the initial azimuthal angle $\phi_0$ as well, even though the eigenstates of the evolved state do depend on $\phi_0$. In summary, the speed of evolution is obviously zero for initial states belonging to the $z$-axis, being them invariant under parallel dephasing; it is furthermore symmetric with respect to the initial azimuthal angle $\phi_0$, and it arises only from the populations of the evolving state when starting from the equatorial $xy$-plane with zero frequency $\omega_0$. 



\begin{figure}[t!]
\includegraphics[width=8.5cm]{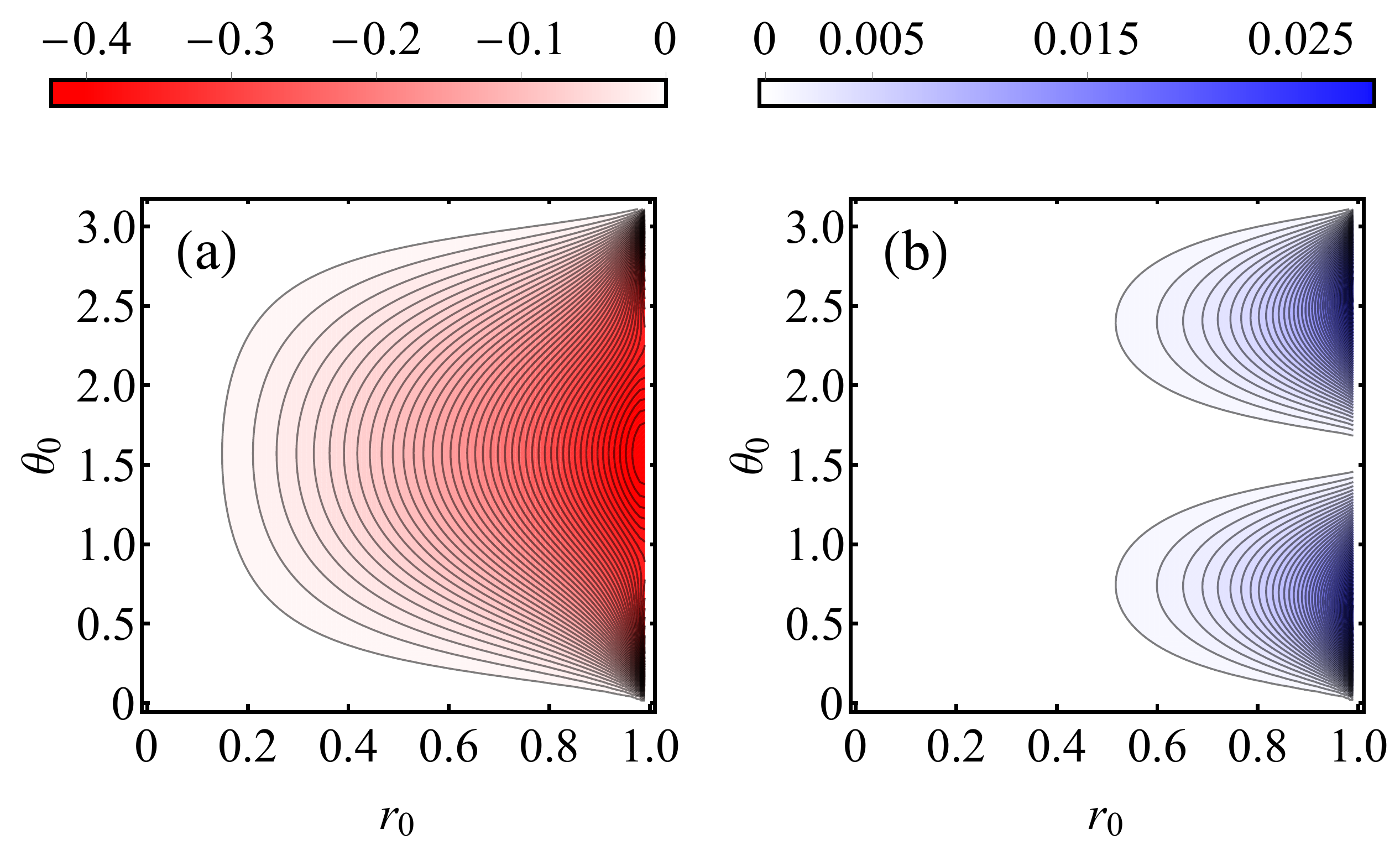}
\caption{(Color online) Contour plot of the difference $\Delta\delta_\gamma \equiv {\delta_\gamma^{QF}} - {\delta_\gamma^{WY}}$ between the tightness parameter corresponding to the quantum Fisher information metric and the one corresponding to the Wigner-Yanase information metric, for the parallel dephasing process as a function of $r_0$ and $\theta_0$, for (\textbf{a}) $\Gamma\tau=10$, $\omega_0=10$, and (\textbf{b}) $\Gamma\tau=10$, $\omega_0=0.1$. The QSL constructed with the Wigner-Yanase skew information is tighter than (respectively looser than) the one constructed with the quantum Fisher information when $\Delta\delta_\gamma > 0$ (resp.~$\Delta\delta_\gamma <0$), as in panel \textbf{b} (resp.~\textbf{a}).}\label{fig:contourplotparalleldephasing}
\end{figure}

In Fig.~\ref{FIG005} we compare the evolution path lengths appearing in the right hand side of Eq.~\eqref{eq:moroimpr003} and corresponding to the three paradigmatic examples of contractive Riemannian metrics: the quantum Fisher information metric, the Wigner-Yanase information metric, and the metric corresponding to the minimal MC function. We consider the only initial parameters that play a role in all the above analysis, i.e. the initial purity $r_0$ and polar angle $\theta_0$, and the dynamical parameter $\beta\equiv\omega_0/\Gamma$, while full details on the computation of all the quantities appearing in Eq.~\eqref{eq:moroimpr003} are deferred to  Appendix~\ref{sec:PD}. First, it can be seen that by fixing the initial purity $r_0$ (respectively, polar angle $\theta_0$), the speed of evolution increases as we increase the initial polar angle $\theta_0$ (respectively, purity $r_0$). In other words, the farther the initial state is from the $z$-axis (the larger is its quantum coherence), the faster the corresponding evolution can be. Second, Fig.~\ref{FIG005}(d) in particular unveils the signature of the populations of the evolved state into the speed of evolution. Indeed, according to Eq.~\eqref{eq:newcohphasedamp001}, the purely quantum contribution ${\mathcal{Q}^f}$ to the metric is equal to zero for $\theta_0 = \pi/2$ and ${\omega_0} = 0$ ($\beta = 0$). Thus, the speed of evolution $g^f$ is described solely by the term ${\mathcal{F}}$ given in Eq.~\eqref{eq:newfisherphasedamp001} and arising only from the populations of the evolved state. In this case, the speed of evolution remains invariant for any contractive Riemannian metric, since ${\mathcal{F}}$ is common to all of them. However, it is still susceptible to changes depending on the purity and time.

Let us now investigate how the QSLs in Eq.~\eqref{eq:moroimpr003} behave by considering the quantum Fisher information metric and the Wigner-Yanase information metric, whose geodesic lengths are known analytically. In the insets of Fig.~\ref{FIG005} we compare the tightness parameter $\delta^f_\gamma$, as defined in Eq.~\eqref{eq:delta}, when considering these two metrics, for a parallel dephasing dynamical evolution. We can see that for $\beta=8$ the dynamics does not saturate the bound for any of the two metrics, although the quantum Fisher information metric provides in general a slightly tighter QSL. On the other hand, when $\beta=0$ and $\theta_0=\pi/2$, we have that the QSL is saturated for both metrics, whereas for $\beta=0$ and $\theta_0=\pi/4$, it is instead the Wigner-Yanase information metric that provides us with a slightly tighter lower bound.

More generally, it is sufficient to compare the difference between the tightness indicators $\delta^{QF}_\gamma - \delta^{WI}_\gamma$ for the two metrics in the whole parameter space of the parallel dephasing model, to identify in which regime each of the two corresponding bounds is the tightest. This analysis is reported in Fig.~\ref{fig:contourplotparalleldephasing}, showing that the Wigner-Yanase information metric does lead in general to a tighter QSL when the frequency $\omega_0$ is sufficiently small. This is in stark contrast with the case of unitary evolutions, discussed in the previous section, and constitutes a first demonstration of the usefulness of our generalized approach to speed limits in quantum dynamics.

\paragraph{Transversal dephasing.}
We now focus on the case of transversal dephasing noise, which lets an initial state $\rho_0$ evolve as $\rho_t =(1/2){\sum_{i,j = 0}^3}{S_{ij}}{\sigma_i}\rho{\sigma_j}$, where $S$ is a $4\times 4$ hermitian matrix whose non-vanishing elements are given by ${S_{00}} = a + b$, ${S_{11}} = d + f$, ${S_{22}} = d - f$, ${S_{33}} = a - b$, ${S_{03}} = ic$ and ${S_{30}} = -ic$, with
\begin{eqnarray}
a &=& \frac{1}{2}(1 + {e^{-u}}) ~,\label{eq:parameterA} \\
b &=& {e^{-u/2}}\cosh\left(\Omega{u}/2\right) ~, \\
c &=& \frac{2\beta\,{e^{-u/2}}}{\Omega}{\sinh\left(\Omega{u}/2\right)} ~, \\
d &=& \frac{1}{2}(1 - {e^{-u}}) ~,\label{eq:parameterD} \\
f &=& \frac{{e^{-u/2}}}{\Omega}{\sinh\left(\Omega{u}/{2}\right)} ~,
\end{eqnarray}
where $u = \Gamma t$, $\Omega = \sqrt{1 - 4{\beta^2}}$ and $\beta = \omega_0/\Gamma$. It is worthwhile noticing that also the transversal dephasing channel is unital, i.e. it leaves the maximally mixed state invariant. This channel has proven to be of fundamental interest within the burgeoning field of noisy quantum metrology, as shown in Chaves \textit{et al.}~\cite{Chaves2013Noisy,Chaves2015PRX}. More precisely, transversal dephasing noise stands as the relevant scenario whereby one can attain a precision in the estimation of the parameter $\omega_0$ that scales superclassically with the number of qubits, even if such noise applies independently to each qubit (while any superclassical advantage is lost in the case of parallel dephasing noise).

By writing the spectral decomposition of the density operator $\rho_t$ we get
\begin{equation}
{\rho_t} = {\sum_{j = \pm}}{p_j}{|{\theta_t},{\phi_t}\rangle_j}{\langle{\theta_t},{\phi_t}|_j} ~,
\end{equation}
where ${p_{\pm}} = (1/2)(1 + {r_0}{\tilde{\xi_t}})$ and
\begin{eqnarray}
{|{\theta_t},{\phi_t}\rangle_{\pm}} =  \ \ \ \ \ \ \ \ \ \ \ \ \ \  \ \ \ \ \ \ \ \ \ \ \ \ \ \ \ \ \ \ \ \ \ \ \ \ \ \ \ \ \ \ \ \ \ \ \ \ \ \ \ \ \ \ \ \ \ \ \ \ \ \ \ \ \ \ \ \ \ \ \ \ \ \ \ \ \ \ \ \ \ \ \ \\
\frac{1}{N_{\pm}}\left[ [(2a - 1)\cos{\theta_0} \pm {\tilde{\xi_t}}]|0\rangle + [(b + ic){e^{i{\phi_0}}} + f{e^{-i{\phi_0}}}]\sin{\theta_0}|1\rangle\right] ~,\nonumber
\end{eqnarray}
with
\begin{eqnarray}
{\tilde{\xi_t}} &=& \sqrt{{(2a - 1)^2}{\cos^2}{\theta_0} + {\tilde{\zeta_t}}{\sin^2}{\theta_0}} ~, \\
\quad {\tilde{\zeta_t}} &=& {b^2} + {c^2} + {f^2} + 2f[b\cos(2{\phi_0}) - c\sin(2{\phi_0})] ~,
\end{eqnarray}
and $N_\pm$ a normalization constant. By putting the above equations into Eqs.~\eqref{eq:mrzcncpetz00090101} and~\eqref{eq:mrzcncpetz00090102}, one obtains expressions which are too cumbersome to be reported here. However, when restricting to the relevant case of an initial plus state (which is an optimal probe state for frequency estimation), i.e. $\rho_0=|+\rangle\langle +|$ with $|+\rangle=(|0\rangle + |1\rangle)/\sqrt{2}$, one obtains the following simple expressions:
\begin{eqnarray}
{\mathcal{F}} &=& \frac{\beta^4}{\Omega^4}\frac{{e^{-u}}{\sinh^2}\left(\Omega{u}/2\right)}{\left(1 - {e^{-u}}\mathcal{G}\right)\mathcal{G}} ~,\\
{\mathcal{Q}^f} &=& \frac{{\beta^2}}{8\mathcal{G}} {e^{-u}}{c^f} ({p_+},{p_-}) ~,
\end{eqnarray}
where
\begin{equation}
\mathcal{G} = \frac{1}{\Omega^2}\left[\cosh\left(\Omega{u}\right) + \Omega\sinh\left(\Omega{u}\right) - 4{\beta^2}\right] ~.
\end{equation}

\begin{figure}[t!]
\centering
\includegraphics[width=8.5cm]{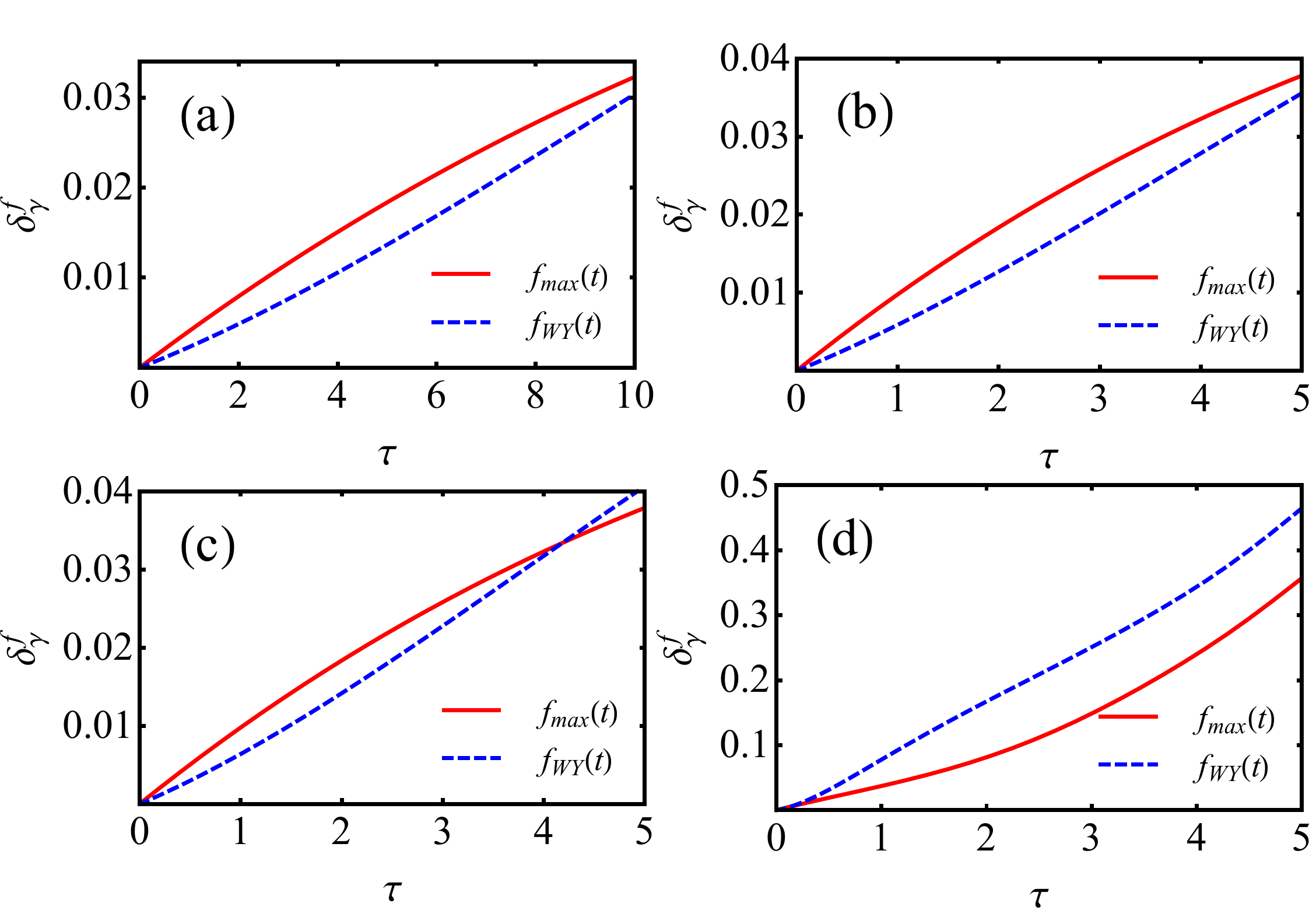}
\caption{(Color online) Plot of the relative difference $\delta^f_\gamma$, Eq.~(\ref{eq:delta}), indicating the tightness of the QSLs (the smaller $\delta^f_\gamma$, the tighter the bounds)  corresponding respectively to the quantum Fisher information (red solid line) and the Wigner-Yanase skew information (blue dashed line), for a qubit initially in the plus state and undergoing a transversal dephasing process with (\textbf{a}) $\Gamma = 0.1$ and $\omega_0 = 0.01$; (\textbf{b}) $\Gamma = 0.25$ and $\omega_0 = 0.01$; (\textbf{c}) $\Gamma = 0.25$ and $\omega_0 = 0.033$; (\textbf{d}) $\Gamma = 1$ and $\omega_0 = 1$.
}
\label{fig:transversaldephasing}
\end{figure}

Let us now analyze the behavior of the QSLs in Eq.~\eqref{eq:moroimpr003} corresponding to the quantum Fisher information metric and the Wigner-Yanase information metric when considering the transversal dephasing dynamics. In Fig.~\ref{fig:transversaldephasing} we can see that, initializing such dynamics with a plus state, it happens that for small enough $\Gamma$ and $\omega_0$ the Wigner-Yanase information provides a QSL which is  tighter (in particular at short times) than the one corresponding to the quantum Fisher information.  One might identify more generally the region of parameters in which this behaviour occurs by studying the trade-off between the respective tightness indicators $\delta^f_\gamma$ for an arbitrary initial state, as in the previous case, although such a study does not add any further insight and is not reported here.

Once more, the present analysis shows that our approach applies straightforwardly to obtain novel, tighter bounds in dynamical cases of interest for quantum technologies, as here corroborated in particular for the metrologically relevant case of transversal dephasing noise.

\subsubsection{Amplitude damping channel}
\label{sec:sectionamplitude}

\begin{figure*}[t]
\centering
\includegraphics[scale=0.43]{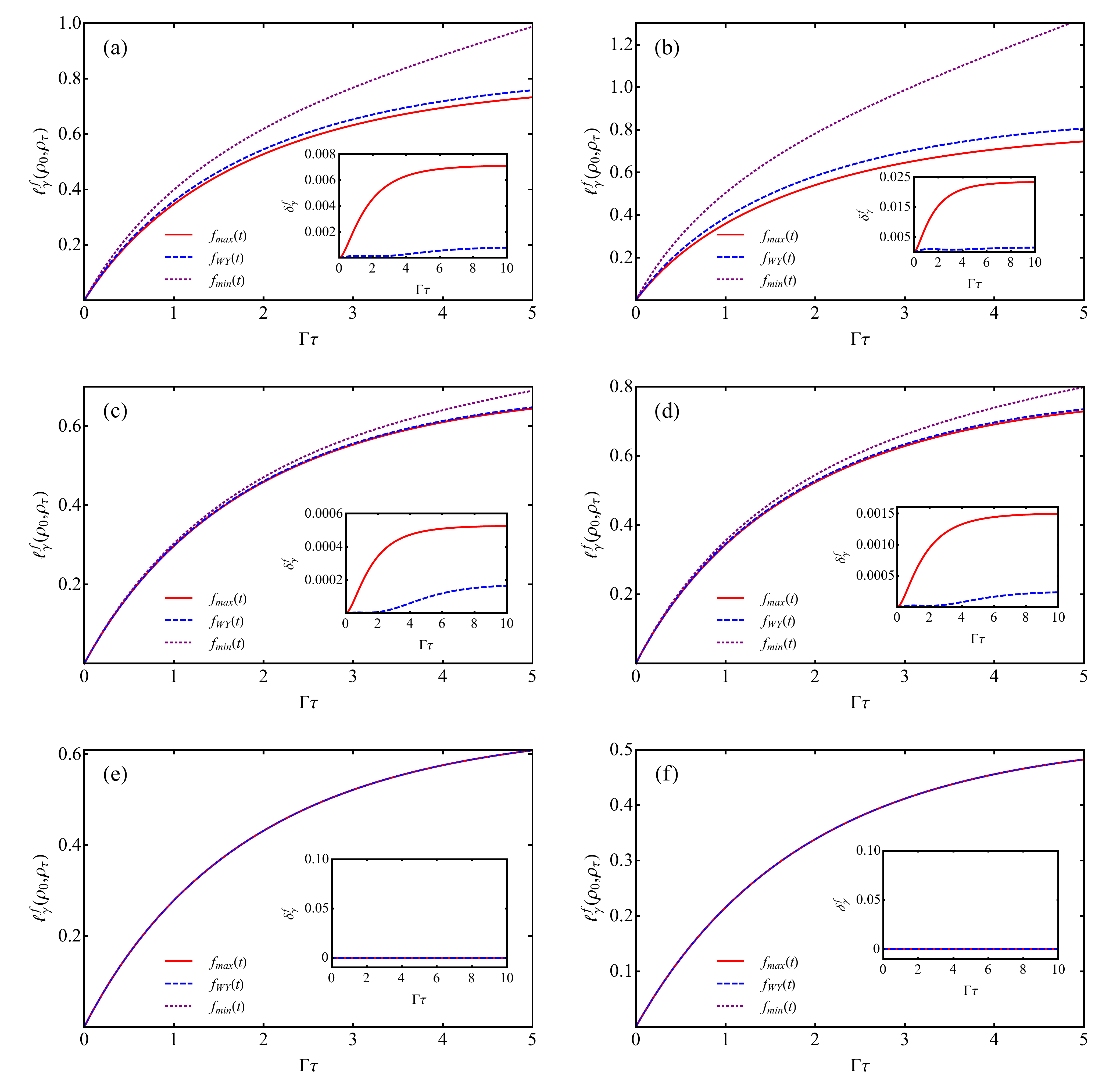}
\caption{(Color online). Evolution path lengths $\ell^f_\gamma$ for amplitude damping processes related to the contractive Riemannian metrics corresponding to the following MC functions: $f_{QF}(t)={f_{max}}(t)$ (red solid line), ${f_{WY}}(t)$ (blue dashed line) and ${f_{min}}(t)$ (purple dotted line) for (\textbf{a}) ${r_0} = 1/2$, $\theta_0 = \pi/2$, (\textbf{b}) ${r_0} = 3/4$, $\theta_0 = \pi/2$, (\textbf{c}) ${r_0} = 1/4$, $\theta_0 = \pi/4$, (\textbf{d}) ${r_0} = 1/4$, $\theta_0 = \pi/2$, (\textbf{e}) ${r_0} = 1/4$, $\theta_0 = 0$ and (\textbf{f}) ${r_0} = 1/2$, $\theta_0 = 0$. The insets in each panel show the relative difference $\delta^f_\gamma$, Eq.~(\ref{eq:delta}), for amplitude damping processes by considering the quantum Fisher information metric (red solid line) and the Wigner-Yanase information metric (blue dashed line); such a relative difference can be regarded as an indicator of the tightness of the bounds (the smaller $\delta^f_\gamma$, the tighter the bounds).}
\label{FIG003}
\end{figure*}

We now consider another canonical model of noise, namely dissipation modelled by an amplitude damping channel acting on a single qubit.
For the amplitude damping channel we have the following Kraus operators
\begin{equation}
 \label{eq:splkrausad0001}
{\tilde{K}_0} = \left(\begin{matrix} 1 & 0 \\ 0 & \sqrt{1 - {\lambda_t}}\end{matrix}\right) ~,\quad
{\tilde{K}_1} = \left(\begin{matrix} 0 & \sqrt{\lambda_t} \\ 0 & 0\end{matrix}\right) ~,
\end{equation}
with $\lambda_t= 1 - {e^{-\Gamma t}}$ and $1/\Gamma$ is the characteristic time of the process~\cite{Nielsen_Chuang_infor_geom}, satisfying only ${\sum_j}{\tilde{K}_j^{\dagger}}{\tilde{K}_j} = \mathbb{I}$ since this channel is not unital. The effect of amplitude damping consists in shrinking the Bloch sphere towards the north pole, or the state $|0\rangle$. In this case, it is easy to verify that the evolved state ${\rho_{t}} = \sum_j \tilde{K}_j \rho_0 {\tilde{K}_j^{\dagger}}=(1/2)(\mathbb{I} + {\vec{r}_{t}}\cdot\vec{\sigma})$ has the following spectral decomposition
\begin{equation}
\label{eq:spectralampltdamp001}
{\rho_{t}} = {\sum_{j=\pm}}\, {p_j}{|\theta_{t},\phi_{t}\rangle_j}{\langle\theta_{t},\phi_{t}|_j} ~,
\end{equation}
where ${p_{\pm}} = \frac{1}{2}(1 \pm \vartheta_t)$ and
\begin{equation}
{|\theta_{t},\phi_{t}\rangle_{\pm}} = \frac{1}{\sqrt{N_{\pm}}}\left[(\varsigma_t\pm \vartheta_t)|0\rangle + {e^{i\phi_0}}{r_0}\sqrt{1-\lambda_t}\sin\theta_0|1\rangle \right] ~,
\end{equation}
with
\begin{eqnarray}
\vartheta_t &=& \sqrt{1 - \zeta_t(1 - {\lambda_t})} ~, \\
\zeta_t &=& 1 - {r_0^2} + {\lambda_t}{(1 - {r_0}\cos\theta_0)^2} ~,\\
\varsigma_t &=& \lambda_t + r_0(1 - {\lambda_t})\cos\theta_0 ~,
\end{eqnarray}
and $N_{\pm}$ a normalization constant. By putting the above equations into Eqs.~\eqref{eq:mrzcncpetz00090101} and~\eqref{eq:mrzcncpetz00090102} one obtains, respectively,
\begin{equation}
{\mathcal{F}} = \frac{[\zeta_t - (1 - {\lambda_t}){(1 - {r_0}\cos\theta_0)^2}]^2}{16{\vartheta_t^2}\zeta_t(1 - {\lambda_t})}
\end{equation}
and
\begin{equation}
\label{eq:ampldampcoh001}
{\mathcal{Q}^f} = \frac{{r_0^2}{\sin^2}\theta_0\, (2-{\varsigma_t})^2\,{c^f}({p_+},{p_-})}{32{\vartheta_t^2}(1 - {\lambda_t})} ~.
\end{equation}
As in the case of the parallel dephasing  channel, both contributions ${\mathcal{F}}$ and ${\mathcal{Q}^f}$ to the speed of evolution $g^{f}$ do not depend on the initial azimuthal angle $\phi_0$. However, contrarily to the parallel dephasing  channel case, here the purely quantum contribution ${\mathcal{Q}^f}$ vanishes only for $\theta_0 = 0,\pi$, whereas the term ${\mathcal{F}}$ vanishes in neither of these cases nor for $\theta_0=\pi/2$, as expected due to the fact that now only the north pole, and not the entire $z$-axis of the Bloch sphere, is left invariant by the dynamics.

In Fig.~\ref{FIG003} we compare the evolution path lengths appearing in the right hand side of Eq.~\eqref{eq:moroimpr003} and corresponding to the usual contractive Riemannian metrics, i.e.~the quantum Fisher information metric, the Wigner-Yanase information metric and the metric corresponding to the minimal MC function, by changing again the initial purity $r_0$ and polar angle $\theta_0$. First, Fig.~\ref{FIG003}(e) and Fig.~\ref{FIG003}(f) exhibit the following behaviour: fixing the initial polar angle $\theta_0 = 0$, the speed of evolution decreases as we increase the initial purity $r_0$. This feature highlights the fact that the north pole of the Bloch sphere is unaffected by the amplitude damping channel. Moreover, according to Eq.~\eqref{eq:ampldampcoh001}, the purely quantum contribution ${\mathcal{Q}^f}$ vanishes identically for $\theta_0 = 0$ and the speed of evolution $g$ and corresponding evolution path length in Eq.~\eqref{eq:moroimpr003} become independent of the choice of the MC function $f$. The nontrivial contribution to the speed of evolution is in this case exclusively due to the term ${\mathcal{F}}$ which depends solely on the populations $p_j$ of the evolved state.

\begin{figure}[t!]
\includegraphics[width=8.5cm]{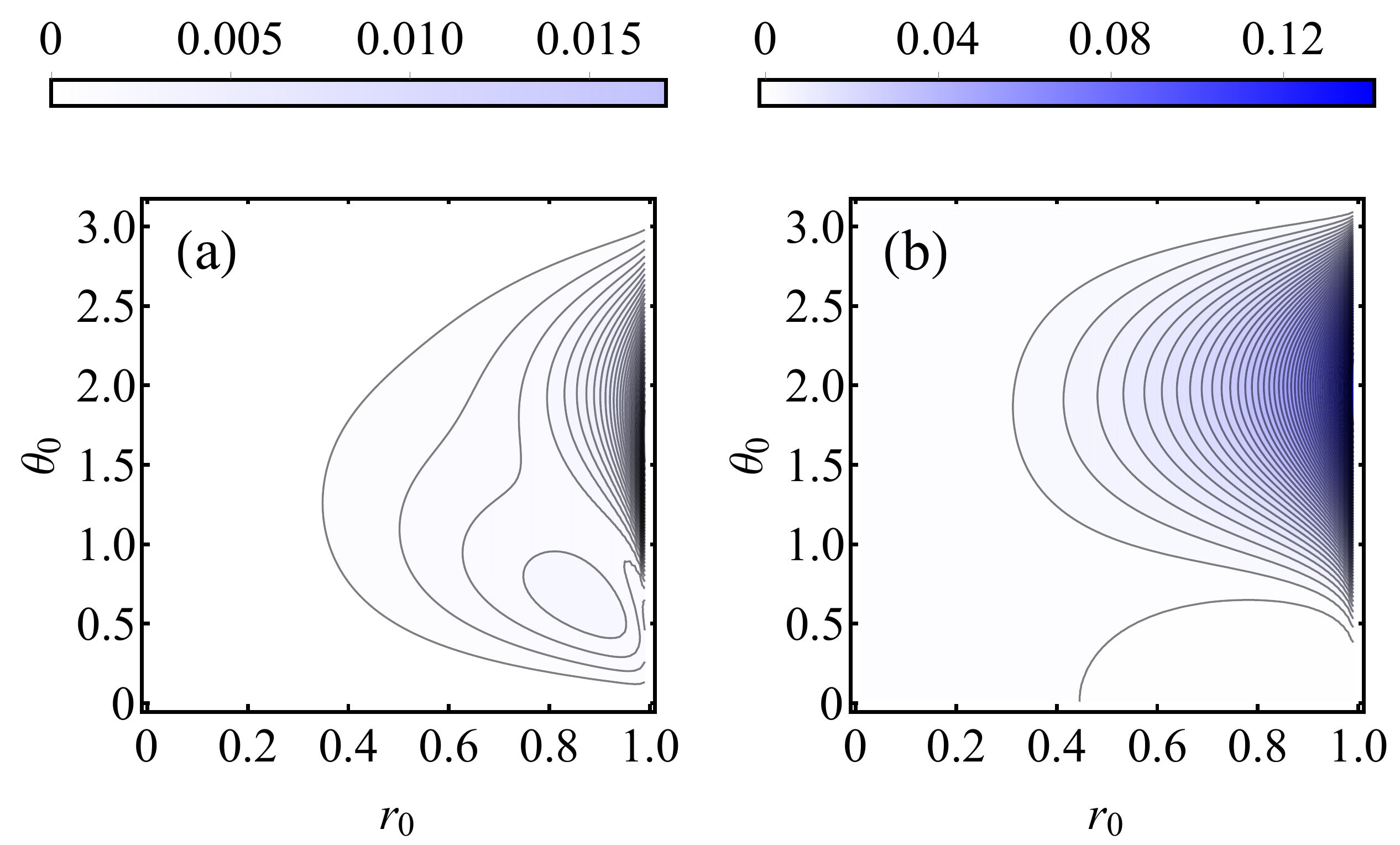}
\caption{
(Color online) Contour plot of (\textbf{a}) the tightness indicator $\delta_\gamma^{WY}$ of the bound specified by the Wigner-Yanase information metric, and  (\textbf{b})
the difference $\Delta\delta_\gamma \equiv {\delta_\gamma^{QF}} - {\delta_\gamma^{WY}}$ between the tightness parameter corresponding to the quantum Fisher information metric and the one corresponding to the Wigner-Yanase information metric, for the amplitude damping process as a function of $r_0$ and $\theta_0$, for $\Gamma\tau=10$. The QSL constructed with the Wigner-Yanase skew information is nearly globally optimal (as $\delta_\gamma^{WY} \simeq 0$) and tighter than the one constructed with the quantum Fisher information (as indicated by  $\Delta\delta_\gamma \geq 0$) in almost the whole parameter space, but for a small region in the bottom-right corner (large $r_0$, small $\theta_0$, i.e.~around the state $|0\rangle$) in which the quantum Fisher information bound is marginally tighter.}\label{fig:contourplotamplitudedamping}
\end{figure}

Let us now analyze the behavior of the QSLs in Eq.~\eqref{eq:moroimpr003} corresponding to the quantum Fisher information metric and the Wigner-Yanase information metric (see Appendix~\ref{sec:AD} for details). In the insets of Fig.~\ref{FIG003} we compare the tightness indicators $\delta^f_\gamma$, as defined in Eq.~\eqref{eq:delta}, when considering these two metrics, for the amplitude damping channel. We can see that in this case the Wigner-Yanase information provides a QSL which is almost saturated (in particular at short times) whereas the quantum Fisher information does not, except in the case of $\theta_0=0$ where they both realize tight bounds. What is more, in Fig.~\ref{fig:contourplotamplitudedamping} one can see that for almost all initial states, except for a small neighbourhood of the north pole (which is the asymptotic state of the amplitude damping channel), it happens that $\delta^{QF} \geq \delta^{WY}$ and $\delta_{WY}\simeq 0$, i.e. the Wigner-Yanase information metric provides us with a definitely tighter (and nearly saturated) QSL than the quantum Fisher information metric.

This reveals another important physical mechanism, distinct from dephasing, in which our generalized analysis leads to significantly tighter bounds than those established in previous literature, in this case clearly demonstrated in almost all the parameter space of relevance. This highlights the power of our general approach to reach beyond the state of the art.

\section{Conclusions}
\label{sec:sectionIV}

Based on the fundamental connection between the geometry of quantum states and their statistical distinguishability, we have exploited the fact that more than one privileged Riemannian metric appear in quantum mechanics in order to introduce a new infinite family of geometric quantum speed limits valid for any physical process, being it unitary or not. Specifically, each bona fide geometric measure of distinguishability gives rise to a different quantum speed limit which is particularly tailored to the case of initial mixed states and such that the contributions of the populations of the evolved state and of the coherences of its time variation are clearly separated. This work provides a comprehensive general framework which incorporates previous approaches to quantum speed limits and leaves room for novel insights.

By investigating paradigmatic examples of unitary and noisy physical processes and of contractive Riemannian metrics, we have seen in fact how the choice of the quantum Fisher information, corresponding to an extremal metric and being ubiquitous in the existing literature, is only a special case which does not always provide the tightest lower bound in the realistic case of open system dynamics. In particular, for parallel and transversal dephasing, as well as amplitude damping dynamics, we defined a tighter quantum speed limit by means of another important but significantly less-studied Riemannian metric, namely the Wigner-Yanase skew information. The bound is useful in practical scenarios of noisy quantum metrology, especially in the case of transversal dephasing \cite{Chaves2013Noisy,Chaves2015PRX}.

Our unifying approach provides a concrete guidance to select the most informative metric in order to derive the tightest bound for some particular dynamics of interest. We have formulated the problem as an optimization of a tightness indicator over all the infinite family of contractive quantum Riemannian metrics. The metric giving rise to the tightest bound is identified as the one whose geodesic is most tailored to the evolution under consideration, see Eq.~(\ref{eq:optimal}). While such a problem can only be solved in restricted form at present, due to the fact that the quantum Fisher information and the Wigner-Yanase skew information are the only two metrics admitting known geodesics, further progress will be achievable in case useful advances on the information geometry for other relevant metrics are recorded in the future.

It is important to remark that the family of speed limits provided in this paper are within the class of MT-like bounds. Following~\cite{Taddeithesis}, it may be possible to implement some adjustments to the adopted unified geometric approach in order to provide a generalized geometric interpretation  to  ML-like speed limits as well. This will be explored in a further study.

Our work readily suggests to explore how the non uniqueness of a contractive Riemannian metric in the quantum state space affects also other scenarios of relevance in quantum information processing. In several of these scenarios, where the quantum Fisher information was adopted and privileged, our approach could lead to a more general investigation based on information geometry.
For example, when considering parameter estimation, one of the paradigmatic tasks of quantum metrology, the inverse of the quantum Fisher information metric sets a lower bound to the mean-square error of any unbiased estimator for the parameters through the quantum Cram\'{e}r-Rao bound~\cite{1994_PhysRevLett_72_3439,2009_Int_J_Quantum_Inform_7_125_Paris}. This work inspires the quest to provide more general bounds on the sensitivity of quantum states to evolutions encoding unknown parameters, based on the infinite hierarchy of quantum Riemannian contractive geometries. It is useful to recall here that the Fisher information-based quantum Cram\'{e}r-Rao bound for single parameter estimation can only be achieved asymptotically in the limit of a large number of probes, and upon performing an optimal measurement given by projection into the eigenbasis of the symmetric logarithmic derivative, which is typically hard to implement in the experimental practice~\cite{2009_Int_J_Quantum_Inform_7_125_Paris}. In the realistic case of a finite number of probes, corrections to the bound provide tighter estimates to the attainable estimation precision; these corrections have been first investigated in Ref.~\cite{Brody1996} for the case of the quantum Fisher information. Motivated by more recent works by Brody~\cite{Brody2011} and~\cite{Brody2012}, in which the Wigner-Yanase skew information has been interpreted rather naturally as the speed of mixed quantum state evolution, and by the analysis of the present work (which includes metrologically relevant settings such as frequency estimation under transversal dephasing \cite{Chaves2013Noisy,Chaves2015PRX}), we believe it is a worthy outlook to investigate finite-size corrections to the Cram\'{e}r-Rao inequality based on the Wigner-Yanase information, in order to determine how tight the bound can be for practical purposes, in particular for the estimation of parameters encoded in open system dynamics. 

Furthermore, within the burgeoning field of quantum thermodynamics, our approach could   provide an infinite class of generalizations of the classical thermodynamical length~{\cite{2007_PhysRevLett_100602}}, originally based on the unique classical contractive Riemannian metric, to the quantum setting. Again in the context of quantum thermodynamics, due to the close connections between geometry and entropy, it might be interesting to investigate the role played by the non uniqueness of a contractive Riemannian geometry on the quantum state space in the existence of many second laws of thermodynamics~\cite{PNAS_2015_Branda_Horodecki_17032015,*2015_Lostaglio_Jennings_Rudolph_NatureCommun_4_6383}. In the study of quantum criticality, within the condensed matter realm, a geometric approach based on the fidelity, i.e. on the quantum Fisher information metric, proved to be fruitful~\cite{2007_PhysRevA_76_062318,*PhysRevLett.99.100603,*PhysRevA.78.042105}. Along the lines of this work, one could apply more general tools associated with any quantum Riemanniann contractive metric, in order to seek for further insights and sharper identification of quantum critical phenomena.

Finally, the general approach presented in this paper to pinpoint the tightest speed limits in quantum evolutions is readily useful for applications to quantum engineering and quantum control. Specifically, the present study allows one to certify that, in a particular implementation, quantum states have been driven at the ultimate speed limit~\cite{2009_PhysRevLett_103_240501} and their evolution cannot be sped up further: this occurs whenever saturation of one of our bounds is demonstrated. As our various examples show, this is not possibly verifiable only by considering the standard bound based on quantum Fisher information. For single-qubit evolutions, we showed that the latter is in fact the tightest for the idealized case of unitary dynamics, while our novel bound based on the Wigner-Yanase skew information can instead be significantly tighter in the most common instances of open dynamics, yielding effectively {\it the} optimal bound (even among all the other unverifiable Riemannian metrics) for amplitude damping dynamics, as certified by a nearly vanishing tightness indicator in such case. Given that the Wigner-Yanase skew information is experimentally accessible \cite{Girolami2014}, one can readily apply our results to current and future demonstrations to benchmark optimality of controlled quantum dynamics in the presence of such ubiquitous noise mechanisms.

In this respect, we would like to point out that an experimental investigation of the main results presented here, for both closed and open system dynamics, can be achieved in particular using a highly controllable Nuclear Magnetic Resonance setup,  with no need for a complete quantum state tomography. In fact, dephasing and amplitude damping are naturally occurring sources of decoherence in such an implementation, and our results can be accessed by means of spin ensemble measurements, which constitute the conventional types of detection in such a technique~\cite{livro_Ivan,PhysRevLett.110.140501}. An experimental investigation as described deserves a study on its own and will be reported elsewhere.

\section{Acknowledgments}

We thank Dorje C. Brody, Tyler J. Volkoff, Frederico Brito, J. Carlos Egues, Paolo Gibilisco and Fumio Hiai for fruitful discussions. The authors would like to acknowledge the financial support from the Brazilian funding agencies CNPq (Grants No.~445516/2014-3, 401230/2014-7, 305086/2013-8, 304955/2013-2 and 443828/2014-8), CAPES (Grant No.~108/2012), the Brazilian National Institute of Science and Technology of Quantum Information (INCT/IQ), and the European Research Council (ERC StG GQCOP, Grant No.~637352).

\appendix

\section{Unitary dynamics}
\label{sec:U}
In this Appendix we prove that the geometric QSL corresponding to the quantum Fisher information metric is tighter than the one corresponding to the Wigner-Yanase information metric, when considering any single-qubit unitary dynamics.

Let us consider a one-qubit state ${\rho_0} = (1/2)(\mathbb{I} + {\vec{r}_0}\cdot{\vec{\sigma}})$, where ${\vec{r}_0} = \{r_0\sin{\theta_0}\cos{\phi_0},r_0\sin{\theta_0}\sin{\phi_0},r_0\cos{\theta_0}\}$ and $\vec{\sigma}$ is the vector of Pauli matrices, which undergoes the generic unitary evolution $\gamma$ specified by ${\rho_t} = {U_t}{\rho_0}{U_t^{\dagger}}$. We want to prove that for any
$\rho_0$ and $U_t$, the following holds (we will drop the subscript $\gamma$ in the remainder of this section for simplicity)
\begin{equation}\label{criterionequbitunitary}
\delta^{QF}\leq \delta^{WY} ~,
\end{equation}
where $\delta^f$ is the tightness indicator corresponding to the contractive Riemannian metric $\textbf{g}^f$ with MC function $f$, as defined in Eq.~\eqref{eq:delta}.
In order to prove the above inequality, we just need to prove that
\begin{equation}\label{criterionequbitunitarysimplified}
\frac{\ell^{QF}(\rho_0,\rho_\tau)}{\ell^{WY}(\rho_0,\rho_\tau)} \leq \frac{\mathcal{L}^{QF}(\rho_0,\rho_\tau)}{\mathcal{L}^{WY}(\rho_0,\rho_\tau)} ~,
\end{equation}
where we denote by $\ell^{f}(\rho_0,\rho_\tau)$ and $\mathcal{L}^f(\rho_0,\rho_\tau)$, respectively, the path length along the given unitary evolution and the geodesic length between initial and final state, $\rho_0$ and $\rho_\tau=U_\tau\rho_0 U_\tau^\dagger$, according to the contractive Riemannian metric $\textbf{g}^f$.

We know that
\begin{equation}
\ell^{f}(\rho_0,\rho_\tau)=\int_0^\tau \sqrt{g^f} dt ~,
\end{equation}
where $g^f = \mathcal{F} + \mathcal{Q}^f$ and
\begin{align}
\mathcal{F} &= \frac{1}{4}\sum_{j=1}^2\frac{1}{p_j}\left(\frac{dp_j}{dt}\right)^2 ~, \\
{\mathcal{Q}^f} &= \frac{1}{2} c^f(p_1,p_2) (p_1-p_2)^2 \mathcal{A}_{12}\mathcal{A}_{21} ~,
\end{align}
with $p_j$ being the eigenvalues of the evolved state $\rho_t$, $\mathcal{A}_{jl}=i \left\langle j\left| \frac{d}{dt} \right|l\right\rangle$ being quantities that depend on the eigenstates $|j\rangle$ of the evolved state $\rho_t$, and finally
\begin{eqnarray}
c^{QF}(p_1,p_2)&=&\frac{2}{p_1+p_2} = 2 ~, \\
c^{WY}(p_1,p_2)&=&\frac{4}{\left(\sqrt{p_1}+\sqrt{p_2}\right)^2} ~.
\end{eqnarray}
Since the eigenvalues ${p_{1,2}} = (1\pm {r_0})/2$ of the unitarily evolving one-qubit state are time independent, it immediately follows that
\begin{eqnarray}
\mathcal{F}&=&0,\\
c^{WY}(p_1,p_2)&=& \frac{4}{1+\sqrt{1-r_0^2}} ~,
\end{eqnarray}
so that
\begin{equation}\label{lefthandside}
\frac{\ell^{QF}(\rho_0,\rho_\tau)}{\ell^{WY}(\rho_0,\rho_\tau)}= \sqrt{\frac{c^{QF}(p_1,p_2)}{c^{WY}(p_1,p_2)}} =  \sqrt{\frac{1+\sqrt{1-r_0^2}}{2}} ~.
\end{equation}
On the other hand, due to Eqs.~\eqref{eq:BuresAngle} and~\eqref{eq:HellingerAngle}, we have that
\begin{equation}
\frac{\mathcal{L}^{QF}(\rho_0,\rho_\tau)}{\mathcal{L}^{WY}(\rho_0,\rho_\tau)} = \frac{\arccos\left[\sqrt{F(\rho_0,\rho_\tau)}\right]}{\arccos\left[ A(\rho_0,\rho_\tau) \right]} ~,
\end{equation}
where the analytical formulae of the fidelity $F(\rho_0,\rho_\tau)$ and the affinity $A(\rho_0,\rho_\tau)$ for any pair of one-qubit states $\rho_0$ and $\rho_\tau$ are the following~\cite{1992_PhysLettA_163_239,1994_JModOpt_41_2315,PhysRevA.91.042330}
\begin{eqnarray}
F(\rho_0,\rho_\tau) &=& \frac{1}{2}\left[1 + {\vec{r}_0}\cdot{\vec{r}_{\tau}} + \sqrt{(1 - {|{\vec{r}_0}|^2})(1 - {|{\vec{r}_{\tau}}|^2})} \, \right],\label{eq:fidelityonequbit}\\
A(\rho_0,\rho_\tau)&=&\frac{1}{4}\left[{\epsilon_0^+}{\epsilon_{\tau}^+} + {\epsilon_0^-}{\epsilon_{\tau}^-}\frac{{\vec{r}_0}\cdot{\vec{r}_{\tau}}}{|{\vec{r}_0}|\,|{\vec{r}_{\tau}}|} \right] ~,\label{eq:affinityonequbit}
\end{eqnarray}
with
\begin{equation}
{\epsilon^{\pm}_a} = \sqrt{1 + {|{\vec{r}_a}|}} \pm \sqrt{1 - {|{\vec{r}_a}|}} ~,
\end{equation}
and $a = 0,\tau$.

Being $\rho_\tau=U_\tau\rho_0U_\tau^\dagger$, it happens that
\begin{eqnarray}
|\vec{r}_0|=|\vec{r}_\tau|={r_0} ~,\\
\vec{r}_0\cdot \vec{r}_\tau = |\vec{r}_0||\vec{r}_\tau|\cos(\varphi_{0,\tau}) = r_0^2\cos(\varphi_{0,\tau}) ~,
\end{eqnarray}
with $\varphi_{0,\tau}$ being the angle between the vectors $\vec{r}_0$ and $\vec{r}_\tau$, so that we simply get
\begin{eqnarray}\label{righthandside}
\frac{\mathcal{L}^{QF}(\rho_0,\rho_\tau)}{\mathcal{L}^{WY}(\rho_0,\rho_\tau)} = \ \ \ \ \ \ \ \ \ \ \ \ \ \ \ \ \ \ \ \ \ \ \ \ \ \ \ \ \ \ \ \ \ \ \ \ \ \ \ \ \ \ \ \ \ \ \ \ \ \ \ \ \ \ \ \ \ \ \\
\frac{\arccos\sqrt{\frac{1}{2}(2-r_0^2 + r_0^2\cos(\varphi_{0,\tau}) )}}{\arccos\left\lbrace\frac{1}{2}\left[\cos(\varphi_{0,\tau})\left(1-\sqrt{1-r_0^2}\right) + 1+\sqrt{1-r_0^2} \, \right]\right\rbrace} ~. \nonumber
\end{eqnarray}

Overall, by collecting Eqs.~\eqref{lefthandside} and~\eqref{righthandside}, we need to prove that
\begin{eqnarray}\label{thefinalcriterion}
\sqrt{\frac{1+\sqrt{1-r_0^2}}{2}} \leq \ \ \ \ \ \ \ \ \ \ \ \ \ \ \ \ \ \ \ \ \ \ \ \ \ \ \ \ \ \ \ \ \ \ \ \ \ \ \ \ \ \ \ \ \ \ \ \ \ \ \ \ \ \ \  \\
\frac{\arccos\sqrt{\frac{1}{2}\left(2-r_0^2 + r_0^2\cos(\varphi_{0,\tau}) \right)}}{\arccos\left\lbrace\frac{1}{2}\left[\cos(\varphi_{0,\tau})\left(1-\sqrt{1-r_0^2}\right) + 1+\sqrt{1-r_0^2} \, \right]\right\rbrace} ~.\nonumber
\end{eqnarray}

\begin{figure}[t!]
\centering
\includegraphics[width=8cm]{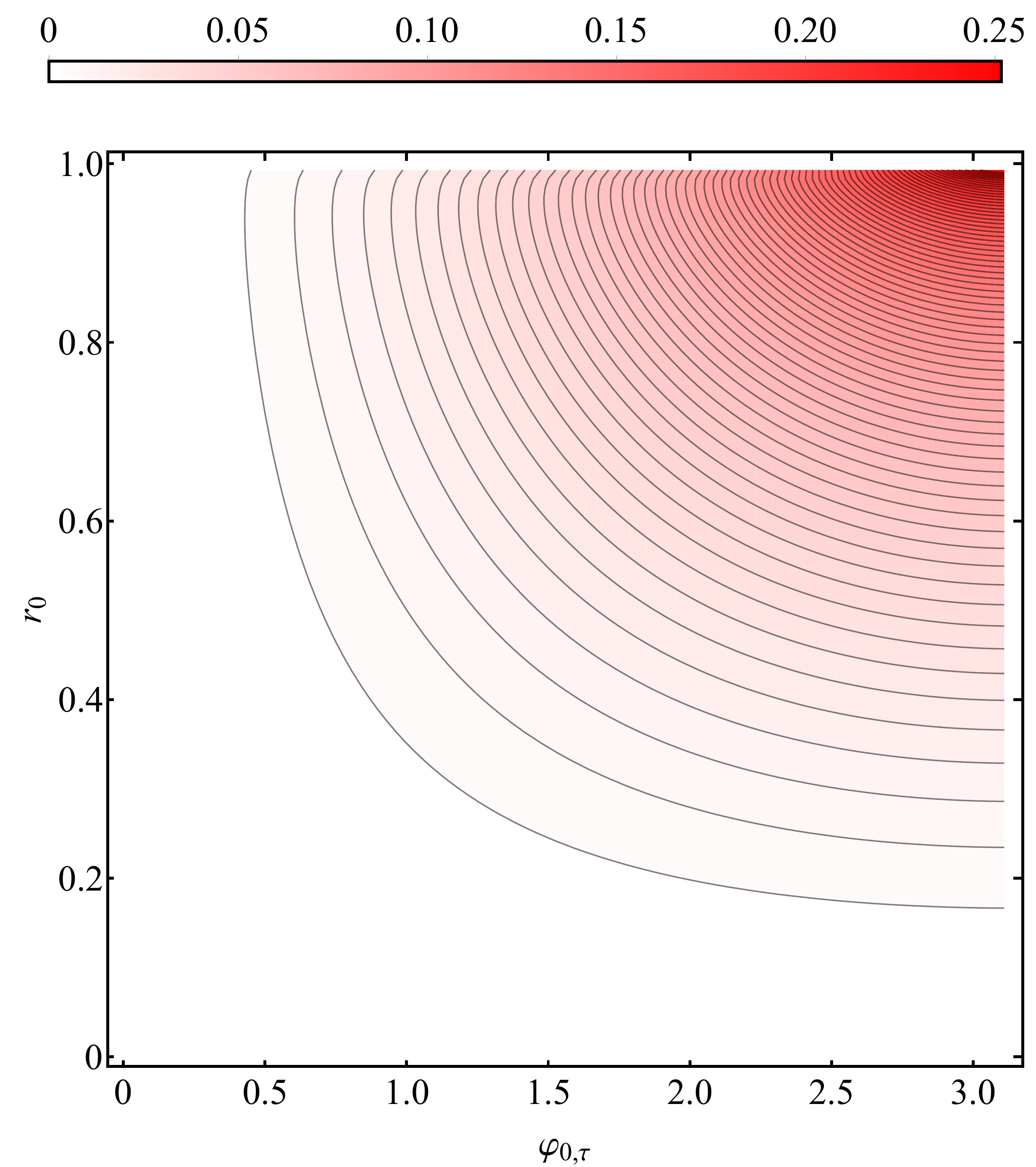}
\caption{The difference between the right hand side and left hand side of Eq.~(\ref{thefinalcriterion}) as a function of $r_0$ and of the relative angle $\varphi_{0,\tau}$.}\label{fig:unitaryonequbitcase}
\end{figure}

The difference between the right hand side and left hand side of the above inequality is represented in Fig.~\ref{fig:unitaryonequbitcase} as a function of $r_0\in[0,1[$ and $\varphi_{0,\tau}\in[0,\pi]$. As it can be easily seen, this difference is always non-negative, i.e. the inequality (\ref{criterionequbitunitarysimplified}) is always satisfied. In particular, this difference is zero when either $r_0=0$ or $\vec{r}_0\cdot \vec{r}_\tau = r_0^2$, i.e. when $\varphi_{0,\tau}=0$, as it can be proved by checking that
\begin{eqnarray}
\lim_{\vec{r}_0\cdot \vec{r}_\tau \rightarrow r_0^2} \frac{\arccos\sqrt{\frac{1}{2}(2-r_0^2 + \vec{r}_0\cdot \vec{r}_\tau)}}{\arccos\left\lbrace\frac{1}{2}\left[\frac{\vec{r}_0\cdot \vec{r}_\tau}{r_0^2}\left(1-\sqrt{1-r_0^2}\right) + 1+\sqrt{1-r_0^2} \, \right]\right\rbrace}= \nonumber \\
\sqrt{\frac{1+\sqrt{1-r_0^2}}{2}} ~.\ \ \ \ \ \ \
\end{eqnarray}

Some remarks are now in order. One the one hand, Eq.~\eqref{lefthandside} can be trivially generalized to any pair of contractive Riemannian metrics $\textbf{g}^f$ and $\textbf{g}^h$ as follows
\begin{equation}
\frac{\ell^{f}(\rho_0,\rho_\tau)}{\ell^{h}(\rho_0,\rho_\tau)} = \sqrt{\frac{c^f(p_1,p_2)}{c^h(p_1,p_2)}} ~.
\end{equation}
However, we also note that this is only true in the one-qubit unitary case. If the dimensionality of the system is higher than 2, then $\mathcal{Q}^f$ becomes a non trivial sum, as defined in Eq.~\eqref{eq:mrzcncpetz00090102}, so that the various time-independent coefficients $c^f(p_j,p_l)$ cannot be extracted from the path integral as we have just done above. On the other hand, Eq.~\eqref{righthandside} seems hard to generalise to other contractive Riemannian metrics, since the analytical expressions of the corresponding geodesic lengths are still unknown. We can thus leave the following conjecture: for any one-qubit unitary dynamics, the quantum Fisher information is the metric which provides the tightest QSL among all contractive Riemannian metrics on the quantum state space, i.e., it is the metric solving the optimization problem in Eq.~(\ref{eq:optimal}). Here we have shown that the conjecture holds when the optimization is restricted to the quantum Fisher information and the Wigner-Yanase skew information metrics only.

\section{Parallel and transversal dephasing}
\label{sec:PD}

In this Appendix we show how to compute all the quantities playing a role in Eq.~\eqref{eq:moroimpr003}, for the parallel and transversal dephasing  dynamics, when considering the quantum Fisher information metric, Wigner-Yanase information metric, and the metric corresponding to the minimal MC function.

Let us start from the case of parallel dephasing, and from the quantum Fisher information metric. Recall that its MC function is the maximal one, i.e. ${f}_{QF}(t) ={f}_{max}(t) = (1 + t)/2$, implying that ${c^f}(x,y) = 2/(x + y)$ and, since $\text{Tr}(\rho_{t}) = {p_+} + {p_-}= 1$,  ${c^f}({p_+},{p_-}) = 2$. It is possible to provide an analytical expression for the path length corresponding to such MC function as follows
\begin{align}
\label{eq:generalTaddei}
{\ell^{QF}_{\gamma}}({\rho_0},{\rho_{\tau}}) &= \frac{1}{2}\sqrt{(1 + {\beta^2})\langle\Delta{\hat{Z}}\rangle}\, \left[ E\left( \arcsin \alpha,{\kappa^2}\right) \right. \nonumber \\ &\left. - E\left( \arcsin\left(\alpha {e^{-\Gamma\tau}}\right),{\kappa^2}\right)\right] ~,
\end{align}
where $E(y,{\kappa^2})=\int_0^y dy' \sqrt{1-\kappa^2 \sin^2 y'}$ is the elliptic integral of second kind, $\beta = {\omega_0}/\Gamma$, ${\kappa^2} = {\beta^2}/(1 + {\beta^2})$ and
\begin{equation}
\label{eq:newvariance001}
\langle\Delta{\hat{Z}}\rangle = 1 - {r_0^2}{\cos^2}{\theta_0} ~,\quad
\alpha = \sqrt{1 - \frac{1 - {r_0^2}}{\langle\Delta\hat{Z}\rangle }} ~,
\end{equation}
with  $\langle\Delta{\hat{Z}}\rangle = {\mbox{Tr}({\rho_t}{\sigma_z^2})} - {\mbox{Tr}({\rho_t}{\sigma_z})^2}$ being the variance of the Pauli matrix $\sigma_z$.
Specifically, when considering the case in which $\omega_0 = 0$, i.e. $\beta = 0$ and $\kappa = 0$, the path length in Eq.~\eqref{eq:generalTaddei} becomes
\begin{equation}
\label{eq:generalTaddei03}
{\ell^{QF}_{\gamma}}({\rho_0},{\rho_{\tau}}) = \frac{1}{2}\sqrt{\langle\Delta{\hat{Z}}\rangle}\, \left[ \arcsin \alpha - \arcsin\left(\alpha {e^{-\Gamma\tau}}\right) \right] ~.
\end{equation}

Furthermore, the geodesic length corresponding to such metric can be readily obtained by using Eqs.~\eqref{eq:BuresAngle} and~\eqref{eq:fidelityonequbit}, with
\begin{eqnarray}\label{eq:newphadampcalc001}
|{\vec{r}_{\tau}}| &=& r_0 \xi_{\tau},\\
{\vec{r}_0}\cdot{\vec{r}_{\tau}} &=& {r_0^2}\left[{\cos^2}\theta_0 + {q_{\tau}}{\sin^2}{\theta_0}\cos({\omega_0}{\tau})\right] ~.\nonumber
\end{eqnarray}

Considering now the Wigner-Yanase information metric, we know that the corresponding MC function is ${f_{WY}}(t) = (1/4){(\sqrt{t} + 1)^2}$ so that ${c^f}(p_+,p_-) = 4/{(\sqrt{p_+} + \sqrt{p_-}\,)^2}$. Hence the path length corresponding to the Wigner-Yanase information metric is given by
\begin{equation}
\label{eq:generalTaddeiWYmetric}
{\ell^{WY}_{\gamma}}({\rho_0},{\rho_{\tau}}) = \frac{1}{2}\sqrt{(1 + {\beta^2})\langle\Delta{\hat{Z}}\rangle}\, {\int_0^{\,\Gamma \tau}}  du\, {\alpha} {e^{-u}}
\, {\Psi_u}({r_0},{\theta_0},\kappa) ~,
\end{equation}
where
\begin{equation}
{\Psi_u}({r_0},{\theta_0},\kappa) = \sqrt{ {\Omega_u}({r_0},{\theta_0},\kappa) +  \frac{1 - {\kappa^2}{\alpha^2}{e^{-2u}}}{1 - {\alpha^2}{e^{-2u}}} }
\end{equation}
and
\begin{equation}
{\Omega_u}({r_0},{\theta_0},\kappa) = \frac{{r_0^2}({{\cos^2}{\theta_0}} + {e^{-2u}}{\kappa^2}{\sin^2}{\theta_0})}{\left[1 + \sqrt{(1 - {r_0^2}{\cos^2}{\theta_0})(1 - {\alpha^2}{e^{-2u}})}\, \right]^2} ~.
\end{equation}

Furthermore, the geodesic length corresponding to such metric can be readily obtained by using Eqs.~\eqref{eq:HellingerAngle},~\eqref{eq:affinityonequbit} and~\eqref{eq:newphadampcalc001}.

Finally, when considering the metric corresponding to the minimal MC function, $f_{min}(t)=2t/(1 + t)$, we have that $c^f(p_+,p_-)=1/(2p_+p_-)$ and so its path length has the following analytical form
\begin{equation}
\label{eq:generalTaddeiharmonicmean}
{\ell^{min}_{\gamma}}({\rho_0},{\rho_{\tau}}) = \frac{1}{2}\sqrt{1 + {\beta^2}}\, \left[ \arcsin \alpha - \arcsin\left(\alpha {e^{-\Gamma\tau}}\right)\right] ~.
\end{equation}
Unfortunately, the geodesic length of the metric corresponding to the minimal MC function is still analytically unknown.

On the other hand, regarding transversal dephasing noise, we are able to compute analytically both the geodesic distances corresponding to the quantum Fisher information and the Wigner-Yanase information metric by using Eqs.~\eqref{eq:BuresAngle},~\eqref{eq:HellingerAngle},~\eqref{eq:fidelityonequbit} and~\eqref{eq:affinityonequbit} with
\begin{eqnarray}
|{\vec{r}_{\tau}}| &=& {r_0}{\tilde{\xi_{\tau}}} \\
{\vec{r}_0}\cdot{\vec{r}_{\tau}} &=& {r_0^2}[(2a - 1){\cos^2}{\theta_0} + [b + f\cos(2{\phi_0})]{\sin^2}{\theta_0}] ~.\nonumber
\end{eqnarray}
Notice that we used the constraint $a + d = 1$, with $a$ and $d$ given respectively in Eqs.~\eqref{eq:parameterA} and~\eqref{eq:parameterD}. The expressions of the path lengths are instead too cumbersome to be reported here in the transversal dephasing case.

\section{Amplitude damping}
\label{sec:AD}

In this Appendix we show how to compute all the quantities playing a role in Eq.~\eqref{eq:moroimpr003}, for the amplitude damping dynamics, when considering the aforementioned three metrics. Similarly to the previous case, we are able to compute analytically both the geodesic distances corresponding to the quantum Fisher information and the Wigner-Yanase information metric by using Eqs.~\eqref{eq:BuresAngle},~\eqref{eq:HellingerAngle},~\eqref{eq:fidelityonequbit} and~\eqref{eq:affinityonequbit} with
\begin{equation}
{\vec{r}_{\tau}}\cdot{\vec{r}_0} = {r_0}\left\{[{\lambda_{\tau}} + (1 - {\lambda_{\tau}}){r_0}\cos{\theta_0}]\cos{\theta_0} + \sqrt{1 - {\lambda_{\tau}}}\,{r_0}{\sin^2}{\theta_0}\right\} ~,
\end{equation}
whereas the geodesic length relative to the minimal MC function is not known analytically.

Now we focus on the evaluation of the path length ${\ell^f_{\gamma}}({\rho_0},{\rho_{\tau}})$ for these metrics. For the quantum Fisher information metric, the analytical expression of the path length is given by
\begin{align}
 \label{eq:lengthqfiad0001}
{\ell^{QF}_{\gamma}}({\rho_0},{\rho_{\tau}}) &= \sqrt{1 - {\varepsilon^2}} \left[E\left(\arcsin{\varpi},{\varepsilon^2}\right) \right. \nonumber \\ &\left. - E\left(\arcsin\left({\varpi}{e^{-\Gamma \tau/2}}\right),{\varepsilon^2}\right)\right] ~,
\end{align}
where
\begin{equation}
{\varepsilon} = \sqrt{\frac{{r_0^2}{\sin^2}{\theta_0}}{2(1 - {r_0}\cos{\theta_0})}} ~,\quad
\varpi = \sqrt{\frac{{1 - {r_0}\cos{\theta_0}}}{2(1 - {\varepsilon^2})}} ~.
\end{equation}


For the Wigner-Yanase information metric, the path length is evaluated as
\begin{equation}
\label{eq:generalTaddeiWYmetricamplitude}
{\ell^{WY}_{\gamma}}({\rho_0},{\rho_{\tau}}) = \frac{1}{2}\sqrt{1 - {\varepsilon^2}}\, {\int_0^{\,\Gamma \tau}}  du\, \varpi {e^{-u/2}}
\, {\tilde{\Psi}_u}({r_0},{\theta_0},\varepsilon) ~,
\end{equation}
where
\begin{equation}
{\tilde{\Psi}_u}({r_0},{\theta_0},\varepsilon) = \sqrt{{\tilde{\Omega}_u}({r_0},{\theta_0},\varepsilon) +  \frac{1 - {\varepsilon^2}{\varpi^2}{e^{-u}}}{1 - {\varpi^2}{e^{-u}}}}
\end{equation}
and
\begin{equation}
{\tilde{\Omega}_u}({r_0},{\theta_0},\varepsilon) = \frac{{\varepsilon^2}\,{[1 + 2(1 - {\varepsilon^2}){\varpi^2}{e^{-u}}]^2}}{{\left[ 1 + 2(1 - {\varepsilon^2})\varpi{e^{-u/2}}\sqrt{1 - {\varpi^2}{e^{-u}}}\, \right]^2}} ~.
\end{equation}

Finally, for the metric corresponding to the minimal MC function, the path length is given by
\begin{equation}
\label{eq:generalTaddeiharmonicmean02}
{\ell^{min}_{\gamma}}({\rho_0},{\rho_{\tau}}) = \Theta({\Phi_{\tau}}) - \Theta({\Phi_0}) ~,
\end{equation}
where
\begin{align}
\Theta(x) &= \arctan\left(\frac{\cos{x}}{\sqrt{{\Delta^2} + {\sin^2}{x}}}\right) + \Delta\, \mbox{arctanh}\, (\cos{x}) + \nonumber \\
&+ \frac{\Delta}{2} \ln\left[1 - \frac{2\cos{x}}{2{\cos^2}(x/2) + \Delta\left(\Delta + \sqrt{{\Delta^2} + {\sin^2}{x}}\, \right)}\right] ~,
\end{align}
${\Phi_u} = \arcsin\left(\varpi{e^{-\Gamma u/2}}\right)$ and ${\Delta^2} = {\varepsilon^2}/[4(1 - {\varepsilon^2})]$.




%

\end{document}